\documentclass[a4paper,10pt]{article}
\usepackage[utf8]{inputenc}
\usepackage{float}
\usepackage{graphicx}
\usepackage{hyperref}
\usepackage{amsmath}
\usepackage{subcaption}
\usepackage[export]{adjustbox}
\usepackage{color}
\usepackage[linesnumbered,ruled]{algorithm2e}
\usepackage{longtable}
\usepackage{authblk}
\usepackage{fullpage}
\usepackage{setspace}
\usepackage{sidecap}
\usepackage{pbox}
\usepackage{cite}
\usepackage[capitalise]{cleveref}
\usepackage{tabularx}
\usepackage{placeins}
\usepackage{array, booktabs, makecell,longtable}
\usepackage{nameref}
\usepackage[utf8]{inputenc}

\usepackage{tabularx}
    \newcolumntype{L}{>{\raggedright\arraybackslash}X}

\DeclareUnicodeCharacter{0394}{\ensuremath{\Delta}}

\makeatletter
\def\BState{\State\hskip-\ALG@thistlm}
\def\arraystretch{1.5}
\makeatother

\title{\bf{Wikipedia and Digital Currencies:\\ Interplay Between Collective Attention \\ and Market Performance}}

\author[a,b]{Abeer ElBahrawy}
\author[c,d]{Laura Alessandretti}

\author[a,e,*]{Andrea Baronchelli}

\affil[a]{{\small City, University of London, Department of Mathematics, London EC1V 0HB, UK }}
\affil[b]{{\small The Alan Turing Institute, British Library, $96$ Euston Road, London NW$12$DB, UK}}
\affil[c]{{\small Centre for Social Data Science, University of Copenhagen, DK-1353 Kgs. K\o benhavn K, Denmark}}
\affil[d]{{\small Technical University of Denmark, DK-2800 Kgs. Lyngby, Denmark}}
\affil[e]{{\small UCL Centre for Blockchain Technologies, University College London, UK}}
\affil[*]{{\small Corresponding author:  Andrea.Baronchelli.1@city.ac.uk}}

\date{}

\begin{document}

\maketitle

\begin{abstract}
The production and consumption of information about Bitcoin and other digital-, or ``crypto-'', currencies have grown together with their market capitalization. However, a systematic investigation of the relationship between online attention and market dynamics, across multiple digital currencies, is still lacking. Here, we quantify the interplay between the attention towards digital currencies in Wikipedia and their market performance. We consider the entire edit history of currency-related pages and their view history from July $2015$. First, we quantify the evolution of the cryptocurrency presence in Wikipedia by analyzing the editorial activity and the network of co-edited pages. We find that a small community of tightly connected editors is responsible for most of the production of information about cryptocurrencies in Wikipedia. Then, we show that a simple trading strategy informed by Wikipedia views performs better, in terms of returns on investment, than baseline strategies for most of the covered period. Our results contribute to the recent literature on the interplay between online information and investment markets, and we anticipate it will be of interest for researchers as well as investors.

\end{abstract}

\section*{Introduction}

The cryptocurrency market grew super-exponentially for more than two years until January 2018, before suffering significant losses in the subsequent months \cite{ElBahrawy170623}. Consequence and driver of this growth is the attention it has progressively attracted from a larger and larger public. In this paper, we quantify the evolution of the production and consumption of information concerning the cryptocurrency market as well as its interplay with the market behavior. Capitalizing on recent results showing that Wikipedia can be used as a proxy for the overall attention on the web \cite{yoshida2015wikipedia}, our analysis relies on data from the popular online encyclopaedia.

The first peer to peer currency system, Bitcoin, was created in $2009$ as a realization of Satoshi Nakamoto novel idea \cite{nakamoto2008bitcoin} of digital currency. The system relies on the Blockchain technology and was built to introduce transparent, anonymous and decentralized digital currency. In the beginning, Bitcoin attracted technology enthusiasts, open source advocates and whoever may need less restrictions on across countries money transfer. Over less than $10$ years, Bitcoin gained popularity and was joined by more than $2,000$ cryptocurrencies \cite{coincap}. Some of these cryptocurrencies (altcoins) are replicas of Bitcoin with small changes in terms of protocols and implementation, while others are adopting entirely different protocols. 

Although cryptocurrencies were first introduced as media of exchange for daily payments \cite{ali2014economics}, they have been increasingly used for speculation \cite{glaser2014bitcoin}. Cryptocurrencies can be traded in online exchange platforms and extensive research has looked at the nature and the main usages of Bitcoin, specifically in hope for some hints on the price drivers \cite{wang2017buzz,kristoufek2015main,ciaian2016economics,guo2018predicting,
gajardo2018does,gandal2016can,elendner2016cross}. Comparisons between cryptocurrencies exchange market and the stock market \cite{ceruleo2014bitcoin,ali2014economics} or fiat currencies \cite{yermack2013bitcoin} have been drawn, in an attempt to rationalize the market and its price movements.

Social media platforms nowadays provide researchers with vast amount of data that can signal public opinions or interests. Since stock markets are highly influenced by the rationale of the investors and their interests, several studies investigated the link between online social signals and stock market prices. Pioneering studies showed how signals from Google trends and Wikipedia \cite{moat2013quantifying,preis2013quantifying} or Twitter sentiment \cite{bollen2011twitter,curme2014quantifying} can help anticipate stock prices. 

This approach has been recently extended to investigate the relationship between social digital traces and the price of Bitcoin  \cite{phillips2018cryptocurrency,dickerson2018algorithmic,colianni2015algorithmic,
garcia2014digital,kristoufek2013bitcoin,phillips2017predicting,kim2016predicting,
stenqvist2017predicting,kim2017bitcoin}, or few top cryptocurrencies \cite{phillips2018cryptocurrency}. While these studies showed the importance of relying on different digital sources, a systematic investigation of multiple cryptocurrencies has been lacking so far. Furthermore, only in few cases \cite{garcia2015social,colianni2015algorithmic,dickerson2018algorithmic}, mostly centred on Bitcoin, the analysis incorporated social media signals into an investment strategy in the spirit of \cite{moat2013quantifying}.

Here, we investigate the interplay between the consumption and production of information in Wikipedia and market indicators. Our analysis focuses on all cryptocurrencies with a page on Wikipedia, from July $2015$ until January $2019$.
The article is organized as follows: In ``\emph{State of the art}", we overview the literature on cryptocurrencies and the online attention towards them; in ``\emph{Data collection and preparation}", we describe the datasets and the pre-processing techniques; in ``\emph{Results}", we present the results of our analysis. Namely, we study the interplay between cryptocurrencies' ``\emph{Wikipedia pages and market properties}"; we study in details the ``\emph{evolution of cryptocurrency pages}"; we investigate ``\emph{the role of editors}" of cryptocurrency pages, and, finally, we explore ``\emph{an investment strategy based on Wikipedia traffic}".

\section*{State of the art}
Two main approaches have been suggested to anticipate Bitcoin and cryptocurrencies prices. The first relies on market indicators only, and uses mostly algorithmic trading and machine learning algorithms to predict prices \cite{madan2015automated,chang2009ensemble,jang2018empirical,alessandretti2018anticipating}. The second relies instead on users' data generated online, including Google search trends, Wikipedia views and Twitter data, to predict and rationalize price fluctuations. Although the relevance of altcoins has been increasing \cite{ElBahrawy170623}, most research has focused on the most notable cryptocurrencies only.

Google search trends, Wikipedia views and Twitter data were found to correlate positively with Bitcoin prices \cite{matta2015bitcoin,kaminski2014nowcasting,colianni2015algorithmic,
garcia2014digital, kristoufek2013bitcoin}. Comments and replies on Bitcoin, Ethereum and Ripple forums \cite{bitcointalk,ethereumforum,xrpchat} were found to anticipate their respective prices \cite{kim2016predicting}. Similar results were obtained considering data from the social news aggregator Reddit for Bitcoin, Litecoin, Ethereum, and Monero \cite{phillips2018mutual,phillips2017predicting}.  In \cite{kristoufek2015main,phillips2018cryptocurrency}, the authors showed positive correlation between multiple online signals and the prices' of Bitcoin, Litecoin, Ethereum, and Monero.

The connection between Bitcoin prices and online social signals has allowed to develop successful trading strategies \cite{garcia2015social,dickerson2018algorithmic,kim2017bitcoin}. In \cite{kim2017bitcoin} the authors used a deep learning algorithm and data from Wikipedia, Google search trends, Bitcoin forum \cite{bitcointalk} and cryptocurrencies news website \cite{coindesk} to anticipate Bitcoin prices.

Research focusing on the nature of community discussions and the activity of contributors is very limited. In \cite{jahani2018scamcoins}, the authors analyzed data from the forum “bitcointalk” \cite{bitcointalk} and showed that there are two clear groups of contributors: Investors, who are driving the market hype, and technology enthusiasts, who are interested in the advancement of the cryptocurrency system.

\section*{Data collection and preparation}\label{sec:data_col_prep}

Wikipedia data was collected through the Wikipedia API \cite{wikipediaAPI} and include the daily number of views and the page edit history of the $38$ cryptocurrencies with a page on Wikipedia (see Supplementary materials, S$1$). 

Page-view data range from July $1$st, $2015$ until January $23$rd, $2019$, since earlier data are not accessible through the API. On the other hand, full editing history is accessible through the API, and includes the content of each edit, the editors, the time of creation and the comments to the edits. Repetitive tasks to maintain pages are often carried by automated tools known as ``bots''. Wikipedia requires bots to have separate accounts and names which include the word ``BOT'', in order to make their edits identifiable. We excluded all edits from bots from our analysis. 

We classified edits into two categories, namely edits with new content and maintenance edits. Maintenance edits aim to keep consensual page content by restoring more accurate old version (reverts) and fighting malicious edits (vandalism). We identified reverts by selecting edits comments containing the word ``rv'' or ``revert'' \cite{kittur2007he}, and by creating an MD$5$ hashing scheme \cite{rivest1992md5} to identify identical files. We created an MD$5$ hash for all edits, and we identified edits sharing the same hash with a previous edit as reverts. Reverts which were made specifically to fight vandalism were identified by selecting edits labeled in their associated comment as ``vandalism'' \cite{kittur2007he}. We considered as new content all edits that were not classified as vandalism nor reverts. 

We also collected data on the activity of the most active editors in other Wikipedia pages. To retrieve this data, we used Xtool \cite{xtool}, a web tool providing general statistics on the editors and their most edited pages.

Market data include daily price, exchange volume and market capitalization of cryptocurrencies, and was collected from the `Coinmarketcap' website \cite{coincap}. The price of a cryptocurrency represents its exchange rate (with USD or Bitcoin, typically) which is determined by the market supply and demand dynamics. The exchange volume is the total trading volume across exchange markets.  The market capitalization is calculated as a product of a cryptocurrency circulating supply (the number of coins available to users) and its price. The market share is the market capitalization of a cryptocurrency normalized by the total market capitalization of the market. Price and market capitalization data is only available since April $28$th, $2013$, while volume data is available since December $27$th, 2013.

The Wikipedia-based investment strategy we implement in this paper can be applied only to ``marginally traded" cryptocurrencies. We compiled a list of $16$ such cryptocurrencies from active exchange platforms including Poloniex and Bitfinex (see Supplementary materials, S$2$). Note that these are also the most widely traded currencies \cite{coincap}. In our analysis, we consider that cryptocurrencies can be traded once their trading volume exceeds $100,000$USD. We excluded days where the reported volume did not lie within $2$ standard deviations from the average trading volume, which are likely due to how market exchanges report their exchange volumes \cite{coincapblog}.

\section*{Results}

\subsection*{Wikipedia pages and market properties}

In this section, we investigate the connection between the attention towards cryptocurrencies registered on Wikipedia and the evolving properties of the market. Wikipedia is the $5^{th}$ most visited website on the Internet \cite{alexa}, attractive to a non-expert audience seeking compact and non-technical information. Previous work has shown that Wikipedia traffic can help predicting stock market prices \cite{moat2013quantifying}.

\begin{figure}[h!]
\centering\includegraphics[scale=0.8]{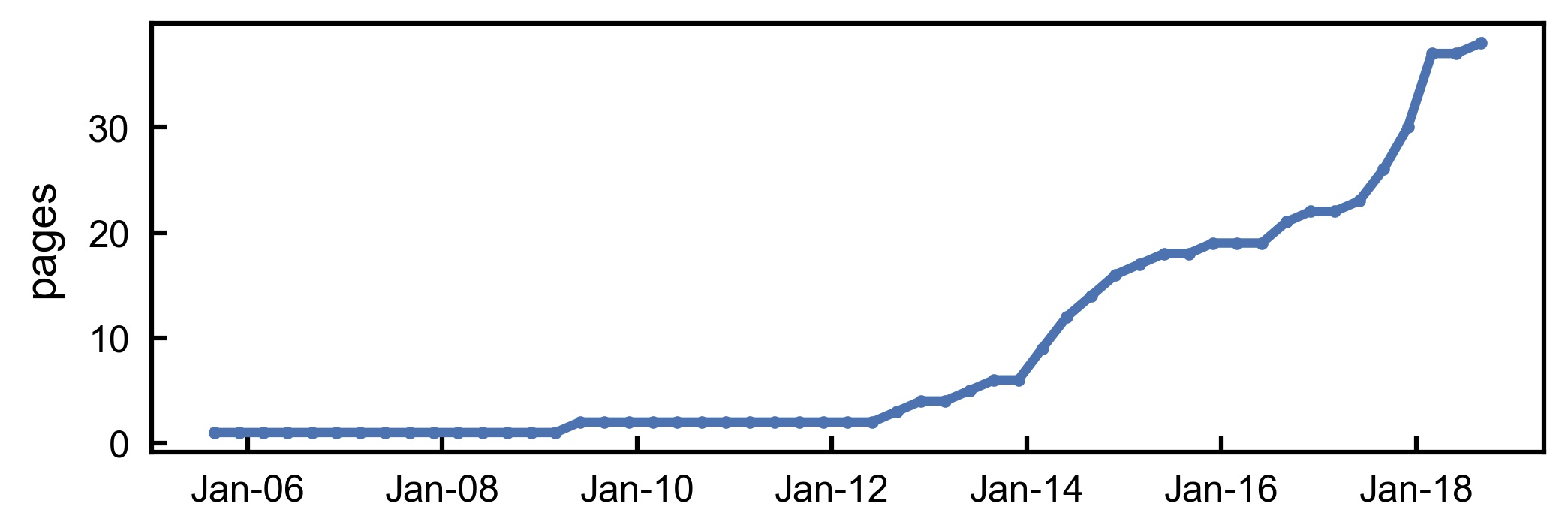}
\caption{\textbf{Cryptocurrencies on Wikipedia.} Evolution in time of the cumulative number of cryptocurrencies with a Wikipedia page.}\label{fig:pages}
\end{figure}

\begin{figure}[H]  
\centering\includegraphics[scale=0.8]{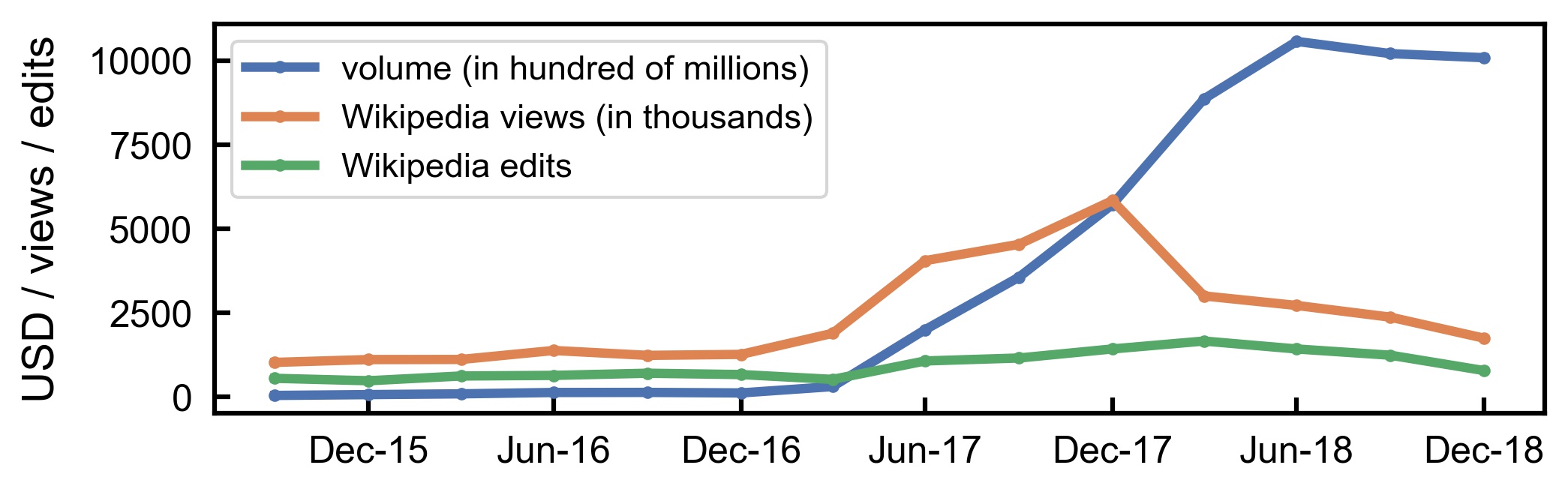}
\caption{\textbf{Market volume and attention to cryptocurrency pages.} The market volume (USD) for all cryptocurrencies with a page in Wikipedia (solid blue line), the total number of views to cryptocurrency pages (solid orange line) and the total number of edits to cryptocurrency pages (solid green line). Values are aggregated using a time window of $3$ months.}\label{fig:views_char}
\end{figure}

The number of cryptocurrency pages on Wikipedia has grown together with their overall market capitalisation. In August $2005$, Ripple became the first cryptocurrency with a page. At that point, it was not identified as a cryptocurrency, but as the idea of a monetary system relying on trust. Bitcoin appeared only in March $2009$, followed by other $36$ currencies (see Figure \ref{fig:pages}).
The number of views received daily by a Wikipedia page is a good proxy for the overall attention on the web \cite{yoshida2015wikipedia}. We find that the number of views to cryptocurrency pages has overall increased from $2015$ until Jan $2018$ (see Figure \ref{fig:views_char}). In $2016$, the $23$ cryptocurrency pages were viewed $\sim 4 \cdot 10^{6}$ times. While in $2017$, $34$ cryptocurrecies pages received $\sim 16 \cdot 10^6$ views. In 2018, the sudden drop in cryptocurrency prices impacted the number of views. The total number of views received by $38$ cryptocurrency pages in $2018$ was $\sim 9 \cdot 10^6$. A second aspect characterizing the evolution in time of Wikipedia pages is their edit history. We find that, on average, pages are more edited than in the past. Cryptocurrency pages ($38$ pages) were edited $\sim 5 \cdot 10^3$ times in $2018$. In $2016$, the $23$ cryptocurrency Wikipedia pages were edited in total $\sim 2 \cdot 10^3$ times (see Figure \ref{fig:views_char}). Bitcoin, in $2016$ was the most viewed cryptocurrency page, with views and edits share of $\sim \%74$ and $\sim \%37$ over all other cryptocurrency pages, respectively. However, these numbers dropped to $\sim \%46$ and $\sim \%16$ in $2018$. The fraction of editors active on Bitcoin's page over all other cryptocurrency pages has also dropped from $\sim 34\%$ in $2016$ to $10\%$ in $2018$. On the other hand, the fraction of views to the $5$ most visited pages compared to all other cryptocurrencies has grown from $\sim \%20$ in $2016$ to $\sim \%27$ in $2018$.  

\begin{figure}[H]
  
\centering\includegraphics[scale=0.8]{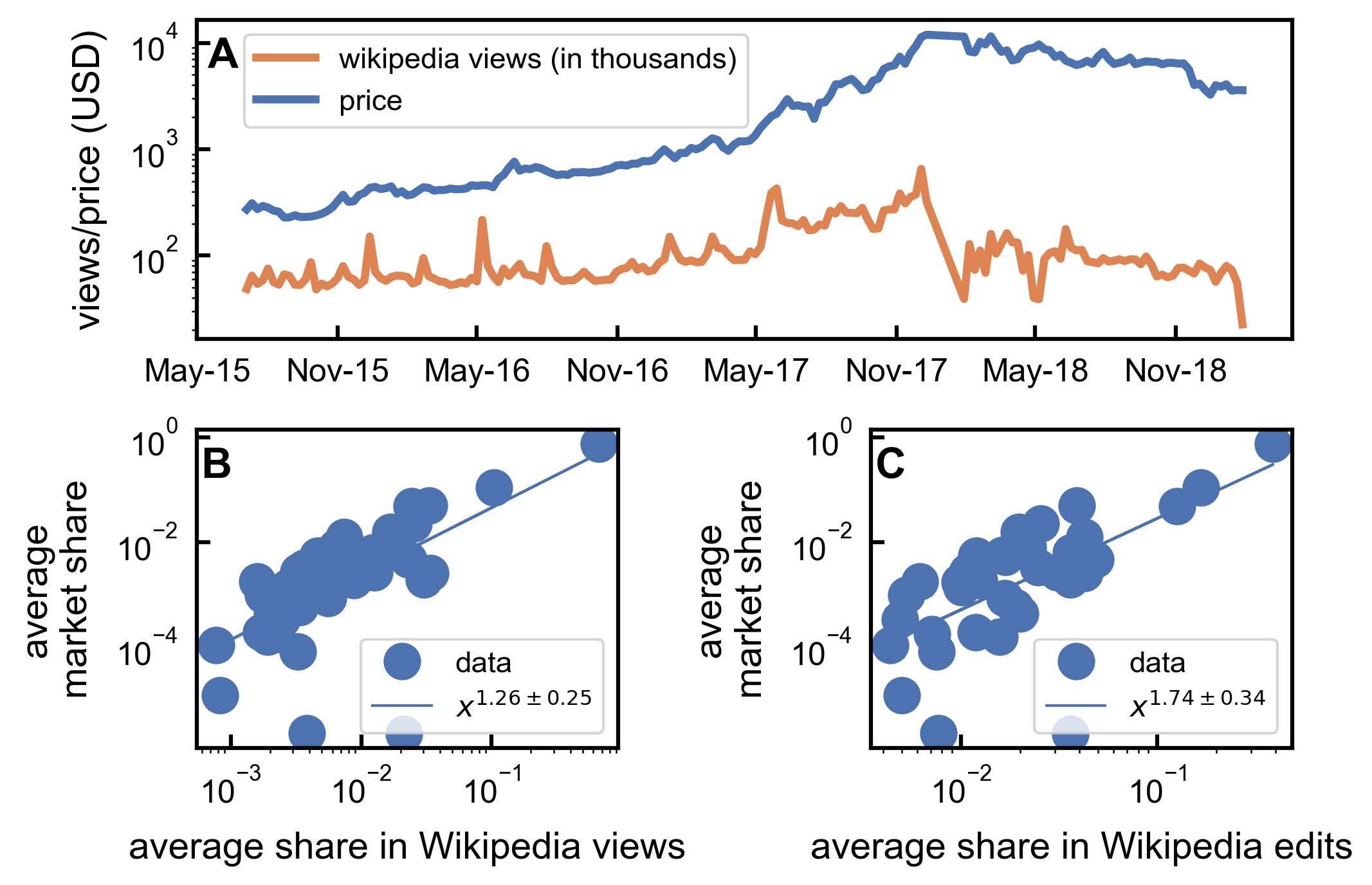}
\caption{\textbf{Correlation between attention on Wikipedia and market performance.} (A) The temporal evolution of price (blue line) and number of Wikipedia views (orange line) for Bitcoin. Averages are computed using a time window of $1$ week. (B) Average market share in $USD$ vs the average Wikipedia views share. Each dot is a different cryptocurrency. (Spearman correlation $\rho=0.71$, $p<10^{-6}$). The solid line represents a power law fit of the data with exponent $\beta = 1.26\pm{0.25}$. (C) Average market share vs the average Wikipedia edits share. (Spearman correlation $\rho=0.68$, $p<10^{-5}$). The solid line represents a power law fit of the data with exponent $\beta = {1.91}\pm{0.34}$}\label{fig:wiki_views_mark_perf}
\end{figure}

Interestingly, Bitcoin's share of the total market capitalization declined during the same period \cite{ElBahrawy170623} suggesting a possible  connection between the properties of the market and the evolution of attention for cryptocurrencies. We find that the daily number of Wikipedia page views and the price of Bitcoin are positively correlated (Pearson correlation $\rho=0.42$, $p<10^{-49}$, see Figure \ref{fig:wiki_views_mark_perf}-A), corroborating the hypothesis of a link between attention on Wikipedia and properties of the market. We further test this hypothesis considering all cryptocurrencies (see Figure \ref{fig:wiki_views_mark_perf}-B) and focusing on other market properties. We find that there is a positive correlation between the average share of views and (i) the average price (Spearman correlation $\rho=0.37$, $p=0.02$), (ii) the average share of volume (Spearman correlation $\rho=0.71$, $p<10^{-7}$), and (iii) the average market share (Spearman correlation $\rho=0.71$, $p<10^{-6}$) of a cryptocurrency. Moreover, these correlations are robust in time (see Figure \ref{fig:ap_wiki_market_corr_time}). 

We also find that the edit history of a currency is connected to the evolution of the market properties (see Figure \ref{fig:wiki_views_mark_perf}-C). We observe a positive correlation between the average fraction of edits and (i) the average price of a given currency (Spearman correlation $\rho=0.36$, $p=0.02$), (ii) the average share of exchange volume for a given currency (Spearman correlation $\rho=0.63$, $p<10^{-5}$) and (iii) its market share (Spearman correlation $\rho=0.68$, $p<10^{-5}$). These correlations are robust in time (see Figure \ref{fig:ap_wiki_market_corr_time}).

\subsection*{Evolution of cryptocurrency pages}

\begin{figure}[hbtp!]
  
\centering\includegraphics[scale=0.8]{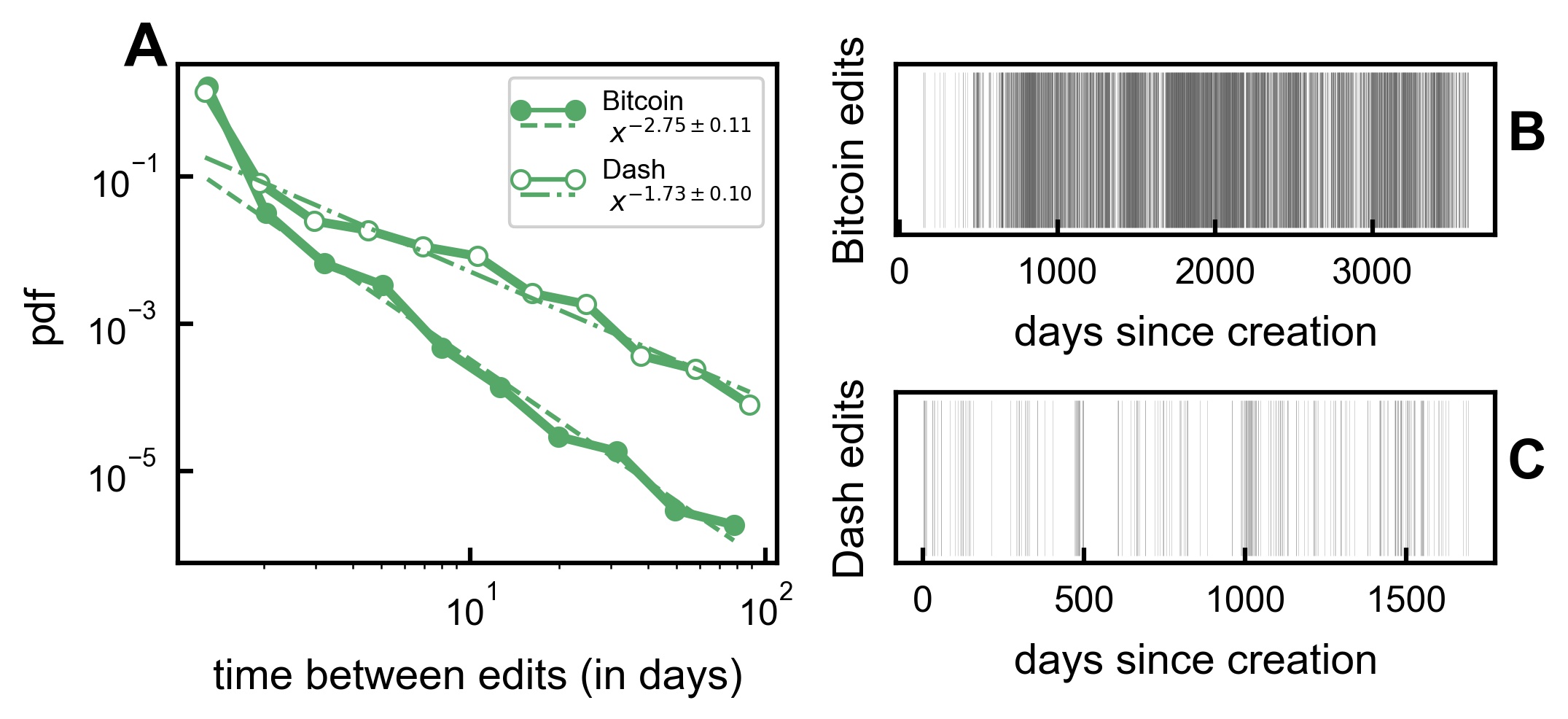}
\caption{\textbf{Example of edit histories.} (A) Distribution of the inter-event time between two consecutive edits for Bitcoin (line with filled circles) and Dash (line with white circles). The dashed line are power-law fits $(P(x)\sim x^{-\beta})$ with exponents $\beta = 2.75$ and $\beta = 1.73$ for Bitcoin and Dash, respectively.
Edits are shown as vertical black line as a function of time for Bitcoin (B) and Dash (C)}\label{fig:edits_pat}
\end{figure}

Frequency of edits and editor diversity are considered reliable indicators of the quality of information included in a Wikipedia page.\cite{stvilia2005assessing}. Cryptocurrency pages differ with respect to their edit history (see Figure \ref{fig:edits_pat}). Some pages, including those of Bitcoin and Ethereum, experience continuous edits throughout their history,  while for other pages, including Dash and Cardano, contributions are intermittent in time, with periods of higher activity followed by calmer ones. For example, the change of the Dash logo in April $2018$ triggered a spike in the number of edits.

\begin{figure}[H]
  
\centering\includegraphics[scale=0.8]{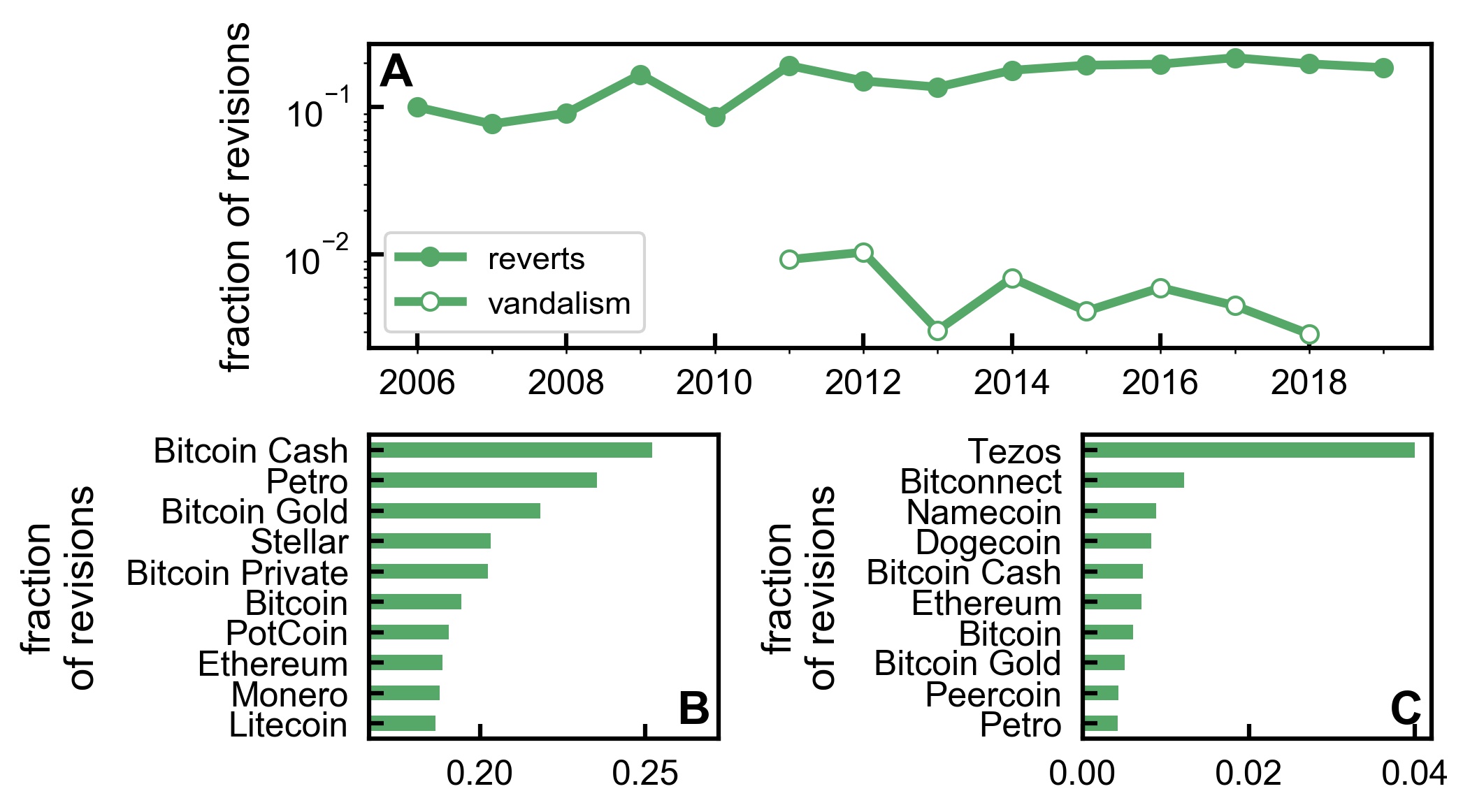}
\caption{\textbf{Reverts and vandalism revisions.} (A) The fraction of ``revert" edits (line with filled circles) and edits reported as vandalism (line with white circles) over time. Values are aggregated using a time-window of one year. B-C) The fraction of reverts (B) and vandalism (C) edits for the top $10$ cryptocurrencies sorted by number of reverts and vandalism edits, respectively.}\label{fig:vand_rev_time}
\end{figure}

The nature of edits changes over a Wikipedia page life. While at the beginning, editors focus largely on new content, as the page ages more efforts are dedicated to fighting vandalism and misinformation (maintenance work) \cite{viegas2004studying,kittur2007he}. We quantify maintenance work by looking at ``reverts", edits that restore a previous version of the page, and at the number of edits reporting vandalism. We find that reverts constitute the $18.2\%$ of all edits, and that, on average, they constitute the $15.4\% \pm 4.3$ of contributions to a cryptocurrency page. The fraction of reverts is stable in time (see Figure \ref{fig:vand_rev_time}-A). Cryptocurrency pages experience higher rates of reverts than an average page in Wikipedia ($8\%$ of the edits at the end of $2016$ \cite{wikivand}), suggesting there is more debate around their content. Only $0.5\%$ of edits were reported as acts of vandalism and their occurrence is constant in time since mid $2011$ (see Figure \ref{fig:vand_rev_time}-A). Well established cryptocurrency pages are less subject to maintenance edits than other pages (see Figure \ref{fig:vand_rev_time}-B and C). Pages of cryptocurrencies forked from Bitcoin such as Bitcoin Cash, Bitcoin Private and Bitcoin Gold were the source of many debates \cite{caffyn2015bitcoin} resulting in a high number of maintenance edits (see Figure \ref{fig:vand_rev_time}-B).

\subsection*{The role of editors}

\begin{figure}[H]
\centering\includegraphics[scale=0.8]{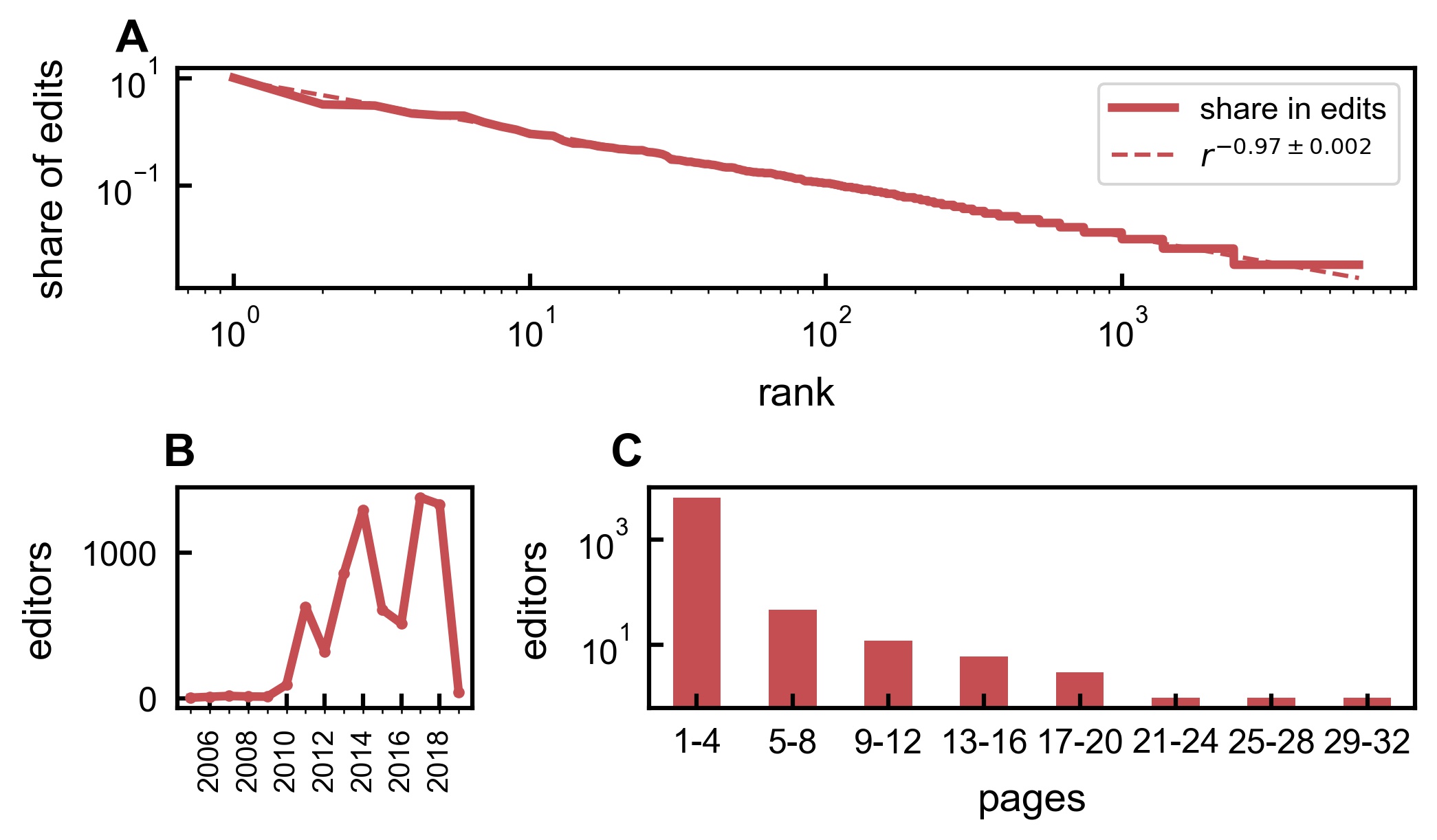}
\caption{\textbf{Uneven distribution of contributions of Wikipedia editors.} (A) The fraction of edits vs the rank $r$ of an editor, over the entire period between $2005$ and $2018$ (red solid line). The dashed line is a power-law fit $(P(r)\sim r^{-\beta})$ with exponent $\beta = 1.01 \pm 0.003$. (B) Number of editors contributing to cryptocurrency pages. Values are aggregated using one year time window. (C) Histogram of editors based on the number of Wikipedia pages they have contributed.}\label{fig:editors_share}
\end{figure}

Our dataset includes $\sim 6170$ editors who contributed $\sim 29,000$ total edits. Although the number of new editors/year fluctuates (see Figure \ref{fig:editors_share}-B, and Appendix \ref{ap:new_pages}), the number of editors has overall increased from $2006$. Only in $2017$, when $10$ new cryptocurrency pages were created, $\sim 1200$ new editors joined. Interestingly, this growth does not characterise all pages on Wikipedia. For example, in \cite{heilman2015wikipedia
}, the authors show that the number of editors in medical related article has been decreasing. 

The editing activity is heterogeneously distributed, as we find by ranking the editors according to the number of edits (see Figure \ref{fig:editors_share}-A). In fact, the relation between rank of an editor, $r$, and fraction of edits can be described by a power law distribution $(P(r)\sim r^{-\beta} )$ where $\beta = 1.01$. This result is in line with what generally observed in Wikipedia \cite{muchnik2013origins}, and consistent across time, with $\beta$ included between $0.84$ and $0.95$ (see Appendix \ref{ap:robust}). In particular, the most active editor alone is responsible for $\sim 10\%$ of the edits (see Appendix \ref{ap:very_active} for more details on the most active editor) and only $\sim 9.6\%$ of the editors ($596$) have edited at least $2$ pages (Figure \ref{fig:editors_share}-C). This group is responsible for $50\%$ of the total number of edits for all cryptocurrency Wikipedia pages.

\begin{figure}[H]
  
\centering\includegraphics[scale=0.8]{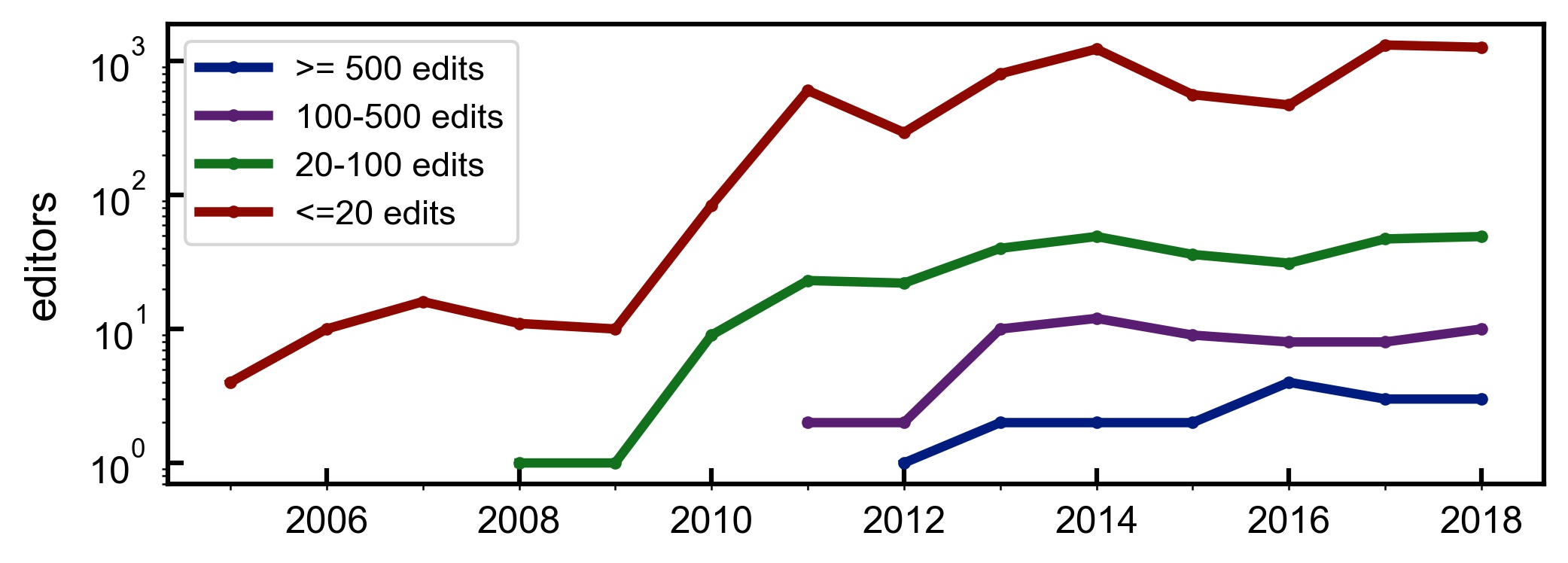}
\caption{\textbf{Active editors per group.} The number of active editors per group from $2005$ until $2018$. Results are computed using a temporal window of one year. Editors are divided in four groups based on their total number of edits: More than $500$ edits (blue line), $100$ to $500$ edits (purple line), $20$ to $100$ edits (green line), less than $20$ edits (red line). Editors were classified according to their total contributions at January $23$rd $2019$,  then traced back.}\label{fig:editors_activity}
\end{figure}

\begin{figure}[H]
  
\centering\includegraphics[scale=0.8]{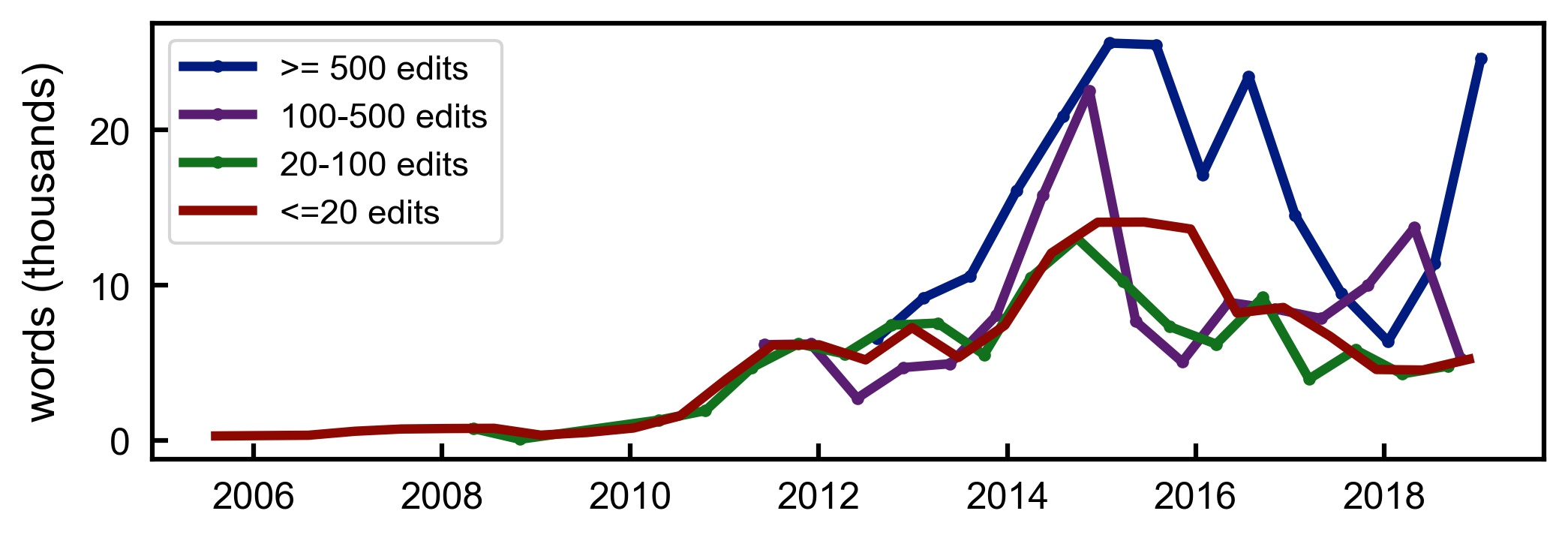}
\caption{\textbf{Activity of editors in different groups.}  Average number of words per editor. All results are computed over a temporal window of $180$ days between August $2005$ and January $2019$. The four lines represent four groups of editors: those who contributed more than $500$ total edits (blue line), $100$ to $500$ edits (purple line), $20$ to $100$ edits (green line), less than $20$ edits (red line). }\label{fig:edits_activity}
\end{figure}

\begin{figure}[H]
  
\centering\includegraphics[scale=0.8]{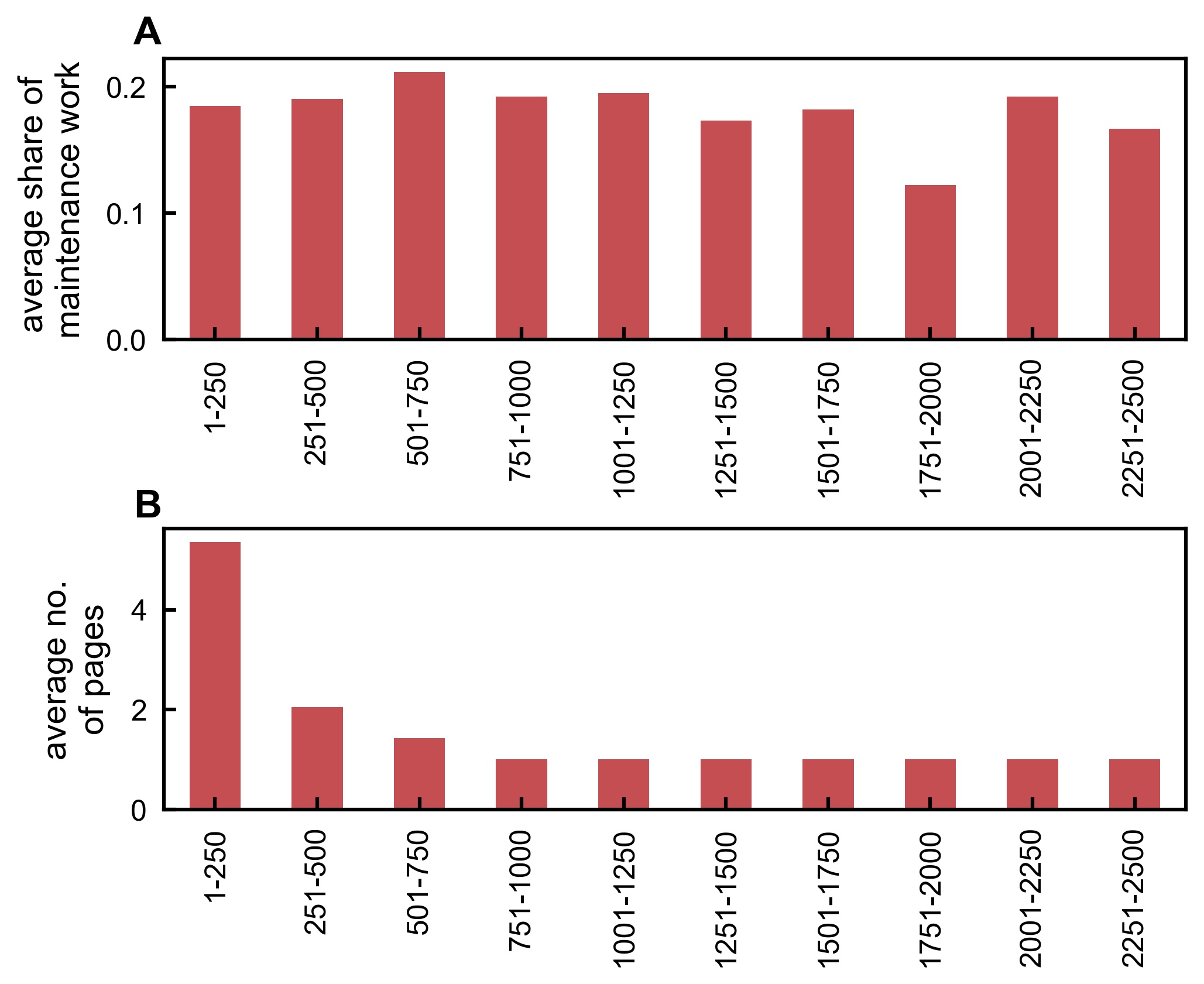}
\caption{\textbf{Focus of editors.} Editors are ranked based on the total number of edits in descending order and grouped based on their rank. (A) Fraction of maintenance edits for each rank group. (B) Average number of contributed pages for each rank group. Only editors with more than one edit are considered.}\label{fig:editors_rank_activity}
\end{figure}

Then, we study the evolution of editors' activity in time. We classify editors into four groups based on their total number of edits at the end of the study, in January $2019$ (see Figure \ref{fig:editors_activity}): Contributors who made more than or equal to $500$ edits ($6$ editors, responsible for $23\%$ of edits), contributors who made $100$ to $500$ edits ($23$ editors, responsible for $15\%$ of edits), contributors who made $20$ to $100$ edits ($142$ editors, responsible for $19\%$ of the edits), editors who made less than $20$ edits ($ 97\%$ of editors, responsible for $43\%$ of the edits). We find that the higher the cumulative activity of a group, the most recently they started editing the pages (see Figure \ref{fig:editors_activity}), in contrast to what is generally observed on Wikipedia \cite{kittur2007power,panciera2009wikipedians}. Note that the group of most active contributors started editing in August $2012$, $3$ years after the creation of Bitcoin's page. Furthermore, Figure \ref{fig:edits_activity} shows that editors with the largest number of edits are responsible for the most extensive contributions in terms of number of edited words. Some of their edits, however, may be for maintenance. By ranking editors in descending order according to their total number of edits across the entire period of study, we find that, for the top $10$ contributors, maintenance edits amount to $20\%$ of their edits. On average, $\sim 18\%$ of the edits written by top $250$ editors are maintenance work (see Figure \ref{fig:editors_rank_activity}-A). This value is consistent among different rank groups. Finally, top ranked editors tend to contribute in more than one page (see Figure \ref{fig:editors_rank_activity}-B), on average $\sim 4$ pages.

\begin{figure}[H]
\centering\includegraphics[scale=0.8]{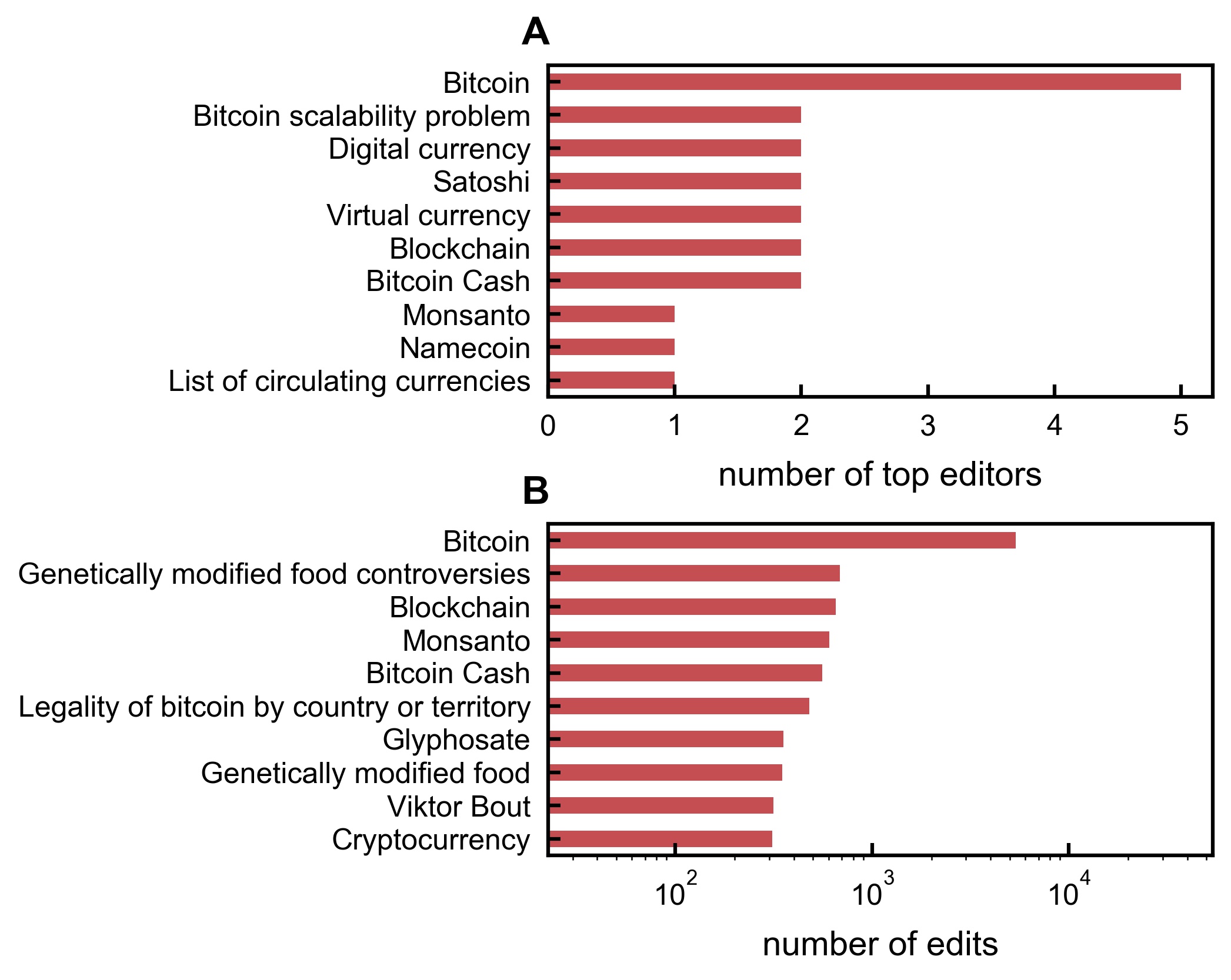}
\caption{\textbf{Activity of the top $6$ cryptocurrency pages editors.} (A) The top $15$ pages by number of editors. The x-axis shows the number of top editors who had this page in their top edited pages. Note that here we consider only the top $10$ pages per editor. (B) The top $15$ pages by number of edits. The x-axis shows the total number of edits per page. Results are obtained for the subset of $6$ most active editors.}\label{fig:top_edit_out}
\end{figure}

To understand the general interests and the specialisation of the top editors of cryptocurrency Wikipedia pages, we focus on a subset of $6$ editors that have contributed at least $500$ edits each. We studied in details their interests by considering their contribution over the entire Wikipedia. Our results show that the main interests of these editors are cryptocurrencies and blockchain (see Figure \ref{fig:top_edit_out}). Results are consistent when we extend the analysis to the top $29$ editors, who are responsible for $37\%$ of the edits. Top editors also contribute in other non-cryptocurrency related pages, however, these pages are less homogeneous and include several different interests such as; genetically modified food, musicians and motor company.

\begin{figure}[H] 
    \centering
    \includegraphics[scale=0.8]{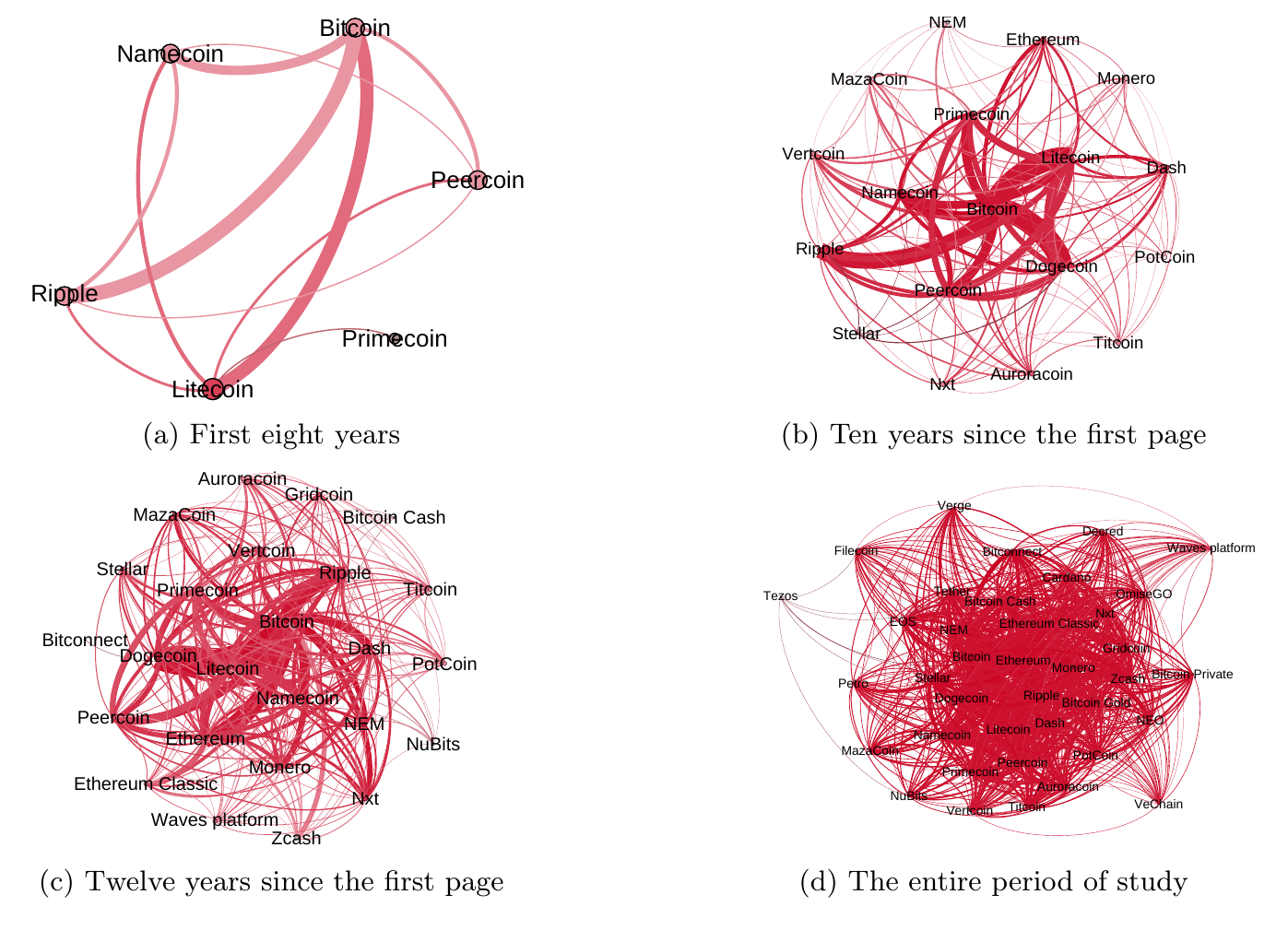} 
  \caption{\textbf{Evolution of the network of cryptocurrency pages.} Nodes represent Wikipedia pages and and edge exist between two nodes if they have at least one common editor. The radius of a node is proportional to the sum of weights of incoming links and the edge thickness is proportional to the edge weight, measured as the number of common editors. The network is aggregated over different period of times: (A) from July $2005$ until July $2013$, (B) from  July $2005$ until July $2015$, (C) from $2005$ until July $2017$, (D) for the entire period of study.}
  \label{fig:network} 
\end{figure}

We further study the network of co-edited Wikipedia pages. We construct an undirected weighted graph, where nodes are Wikipedia pages, an edge exists between two nodes if they have at least one common editor, and link weights correspond to the number of common editors. By the end of July $2014$, the network had $13$ nodes (see Figure \ref{fig:network}-B) and the average node weighted degree was $\langle s \rangle=78.3$ with a total of $2691$ editors.  The weighted degree was heterogeneously distributed: Bitcoin had the largest strength, $s_{BTC}=207$, while recently introduced nodes (Dash, Auroracoin and Nxt) had the lowest weighted degree. These properties have persisted in time (see  Figure \ref{fig:network}-C and \ref{fig:network}-D) and a cryptocurrency page age is positively correlated with its network weighted degree (Pearson correlation $\rho=0.40$, $p=0.015$, see Appendix \ref{fig:ap_s_t}). Bitcoin has the highest degree centrality throughout the entire period considered (see Appendix \ref{fig:ap_cent_years}).

\begin{figure}[H]
  
\centering\includegraphics[scale=0.8]{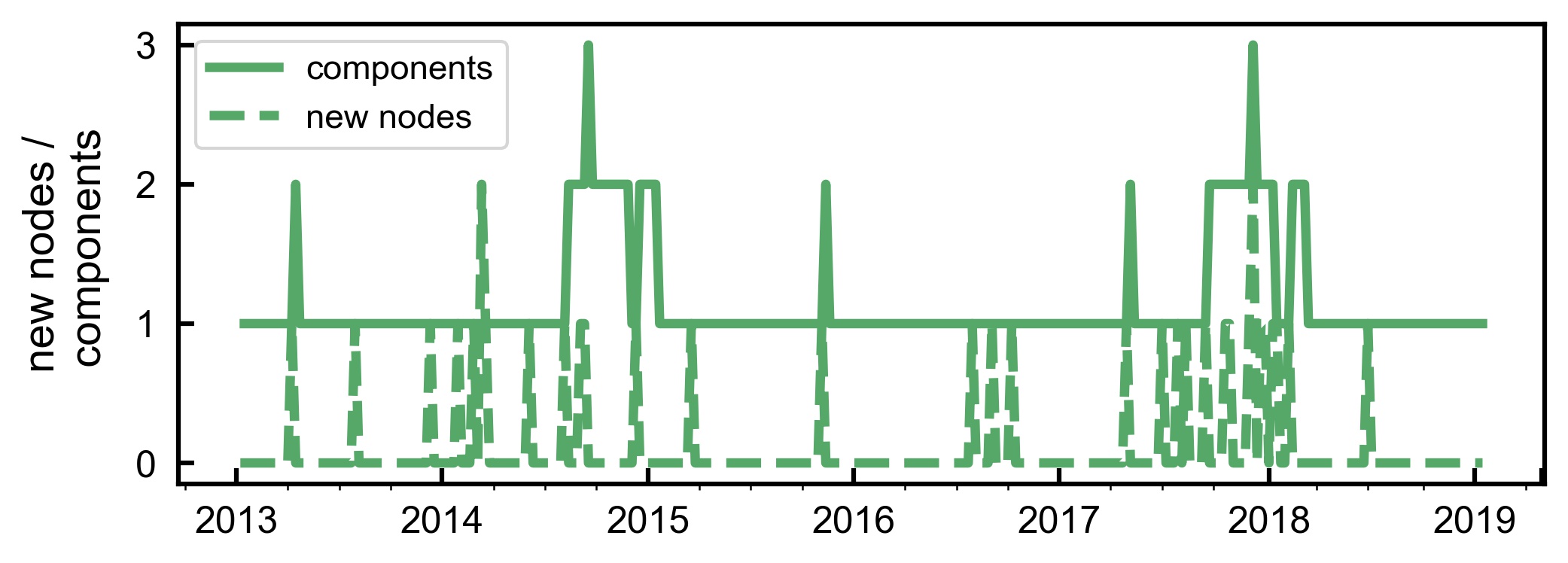}
\caption{\textbf{Short-term dynamics of the Wikipedia network evolution.} Cumulative number of new nodes (dashed line) and total number of network components (solid line). Valurs are aggregted using a $1$ week time window.}\label{fig:com_node_change}
\end{figure}

A giant component (see Figure \ref{fig:network}) emerges in the network, implying each node is connected to all other nodes when we analyse its evolution under large time-windows ($\sim$ years). Instead, if weekly time windows are considered, we find that the network is disconnected (see Figure \ref{fig:com_node_change}). Typically, new pages are created by new editors. On average, new pages connect to the giant component within $5.2$ weeks from creation (see Figure \ref{fig:com_node_change}), in most cases thanks to experienced editors who contribute the newly created page.

\subsection*{An investment strategy based on Wikipedia attention}

The demonstrated connection between the properties of the cryptocurrency market and traffic on Wikipedia suggests the latter could help informing a successful investment strategy. We investigate this possibility by testing a Wikipedia-based strategy similar to the one proposed in \cite{moat2013quantifying,preis2013quantifying} for stock markets investments.

For a given page and a given day $t$, the Wikipedia investment strategy relies on the difference $\Delta n (t)=v(t) - v (t-1)$ between the number of page views $v(t)$ at day $t$ and the number of views $v(t-1)$ at $t-1$. According to the strategy, if $\Delta n(t) > 0$, the investor sells the asset (at price $p(t+1)$) at time $t+1$ and then she buys at time $t+2$ (at price $p(t+2)$). This trading position is formally known as short position. On the other hand, if $\Delta n(t) \le 0$ the investor  buys at time $t+1$ (at price $p(t+1)$) and sells at time $t+2$ (at price $p(t+2)$), which is known as long position. The intuition behind the strategy is that if attention and information gathering has been rising, prices will drop, and vice-versa \cite{moat2013quantifying,tversky1991loss}. We consider Wikipedia views rather than edits, since the latter do not vary on a daily basis (the average time between edits is $10.12$ days). Considering a longer period would overlook the cryptocurrencies' price volatility \cite{brauneis2018price}.

We also consider two baseline strategies. 
The first is based on the price difference $\Delta p (t)=p(t) - p(t-1)$ rather than the page views difference $\Delta n (t)$ \cite{alessandretti2018anticipating}. In all other aspects, it is identical to the Wikipedia-based strategy. This will allow us to test which indicator (price or Wikipedia page views) has better predictive capabilities under the same conditions. The rationale behind the first baseline strategy is that if the price has been rising, a drop will follow, and vice-versa. As a second baseline, we choose a random strategy, where, at every time $t$, one chooses either to buy or to sell an asset with $50\%$ probability \cite{moat2013quantifying}.

The performance of the different strategies is assessed by computing the cumulative return $R$, defiend as the summation of log-returns obtained under the proposed strategies. When $\Delta n(t)>0$ the log-return is computed as $log(p(t+1)) - log(p(t + 2))$, while, in the opposite case, the log-return is $log(p(t + 2)) - log(p(t+1))$. The use of the log returns is motivated by the ease of calculation of the short and long positions and since we are considering multi-period returns \cite{hudson2015calculating}. 

We test the Wikipedia-based strategy against the baselines for the $17$ cryptocurrencies that have a Wikipedia page and can be marginally traded (see list of exchanges with margin trading support in Appendix  \ref{margin_trade}). Margin trading is a practice of borrowing fund from a broker to trade financial assets, that rely on selling assets one does not yet own. 

\begin{figure}[h!]
  
\centering\includegraphics[scale=0.8]{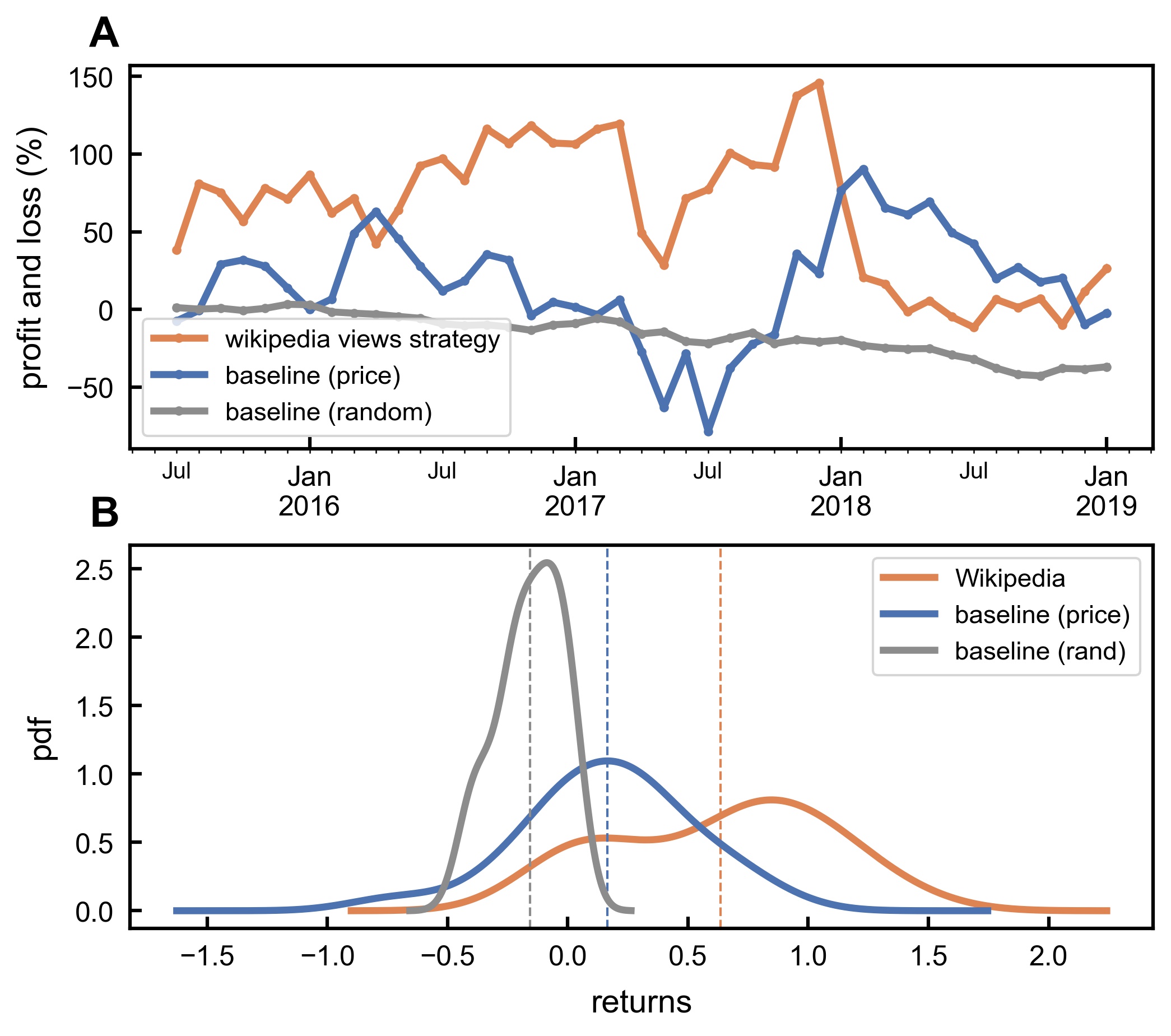}
\caption{\textbf{The Wikipedia based investment strategy outperforms the baseline.} \textbf{(A)} The cumulative return obtained using three investment strategies: the Wikipedia-based strategy (orange line) the baseline strategy based on prices (blue line) and the random strategy (grey line). \textbf{(B)} The distributions of the daily returns obtained using the Wikipedia-based strategy (orange line), the baseline strategy based on prices (blue line) and the random strategy (grey line). The average returns are $\langle r _{w} \rangle = 0.62 \pm   0.42$ (dashed orange line), $\langle r _{p} \rangle = 0.16 \pm 0.36$ (dashed blue line), $\langle r _{r} \rangle = -0.15 \pm 0.13$ (dashed grey line) for the Wikipedia-based strategy, the price based baseline, and the random strategy, respectively. Data is displayed using a kernel density estimate, with a Gaussian kernel and bandwidth calculated using Silverman's rule of thumb. Data for the random strategy is obtained from $1000$ independent realizations. All results are shown for investments between July $2015$ and January $2019$ for all cryptocurrencies which can be traded marginally combined. }\label{fig:wiki_str_ret}
\end{figure}

We find that the Wikipedia based strategy outperforms both baseline strategies, when one considers the period between July $2015$ and January $2018$ (see Figure\ref{fig:wiki_str_ret}-A). On average, the return obtained following the Wikipedia based strategy is  $\langle r _{w} \rangle = 0.62 \pm   0.42$, while the average return obtained random strategy, which is $\langle r _{r} \rangle = -0.15 \pm 0.13$. (see Figure\ref{fig:wiki_str_ret}-B). The distributions of returns obtained under the two strategies are significantly different under Kolomogorov-Smirnov test, with $p \ll 0.05$. The price baseline strategy produces lower mean returns compared to the Wikipedia strategy ($\langle r _{p} \rangle = 0.16 \pm 0.36$).

\begin{figure}[h!] 
\centering\includegraphics[scale=0.8]{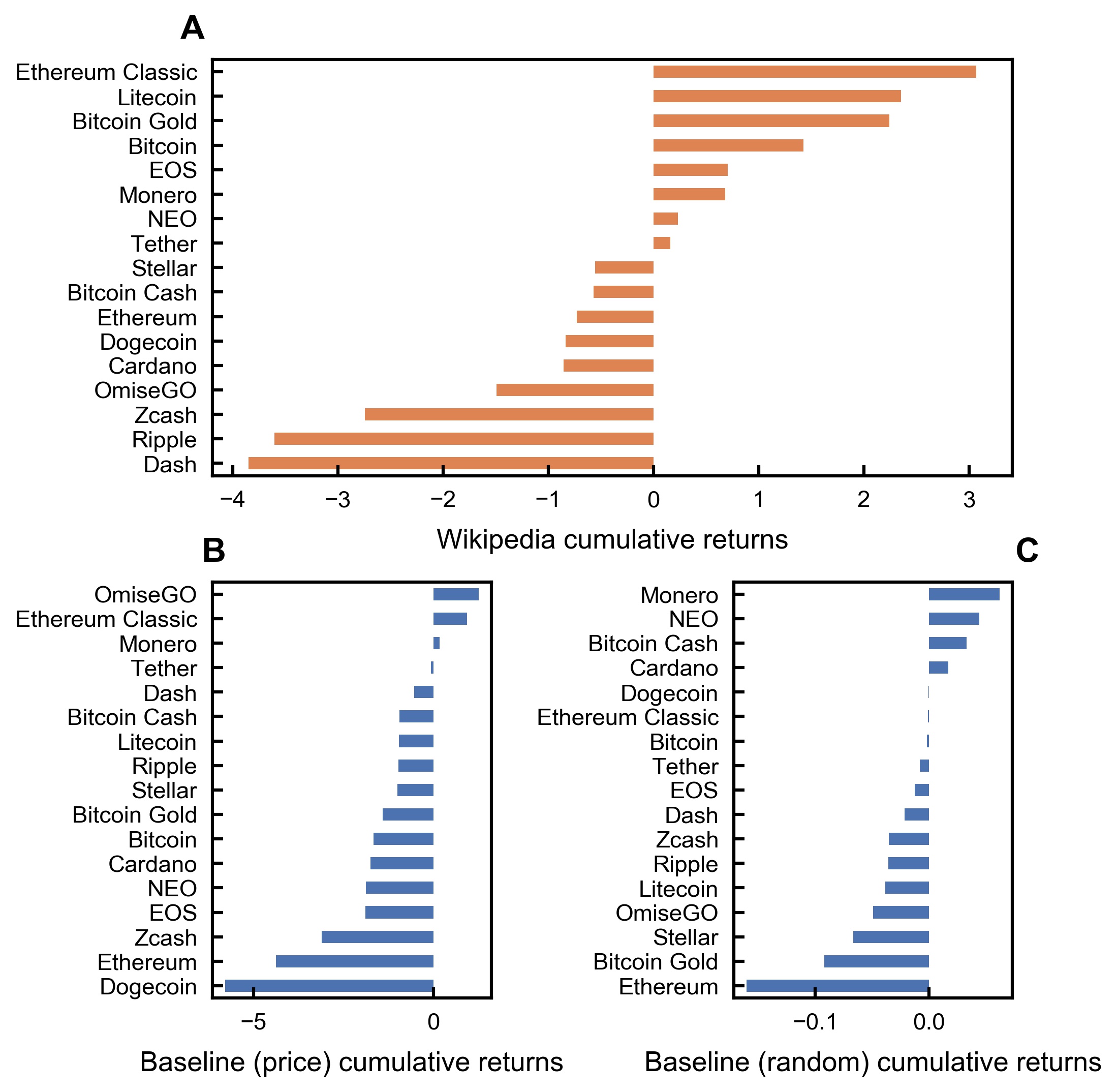}
\caption{\textbf{Performance of the strategies for different cryptocurrencies.} The cumulative returns along the whole period of investment, following the Wikipedia based strategy \textbf{(A)} the price-based baseline strategy \textbf{(B)} and the random strategy \textbf{(C)} for the $16$ cryptocurrencies considered.}\label{fig:crypt_perf}
\end{figure}

A closer inspection shows that there are consistent differences between cryptocurrencies, with respect to the average return (see Figure \ref{fig:crypt_perf}), with some even yielding overall negative returns. The Wikipedia-based strategy yields a positive cumulative returns of $\sim 300\%$ for Ethereum Classic, but for other currencies, including Ripple and Ethereum, investing based on Wikipedia leads to negative returns.

The observed differences could be potentially explained by the correlation between changes in daily price and in Wikipedia views. Instead, we observe that, although the Wikipedia-based strategy works well for Bitcoin but not for Dash, for both currencies there is a positive correlation between daily change in price and Wikipedia views of $0.1$ and $0.18$ respectively (see Figure  \ref{fig:ap_wiki_views_mark_perf_crypt}). However, our proposed strategy does not simply map to buying a cryptocurrency when its Wikipedia page views increases. In order to gain positive returns using our proposed strategy, an increase of the number of views at time $t$, should be followed by an increase in price in the next day $t+1$ and a decrease of the price in the day after $t+2$. Positive returns will also occur in case of a decrease in the number of views at time $t$ if it was followed by a decrease in the price at time $t+1$ and an increase in price at time $t+2$. 

\begin{figure}[h!]
\centering\includegraphics[scale=0.8]{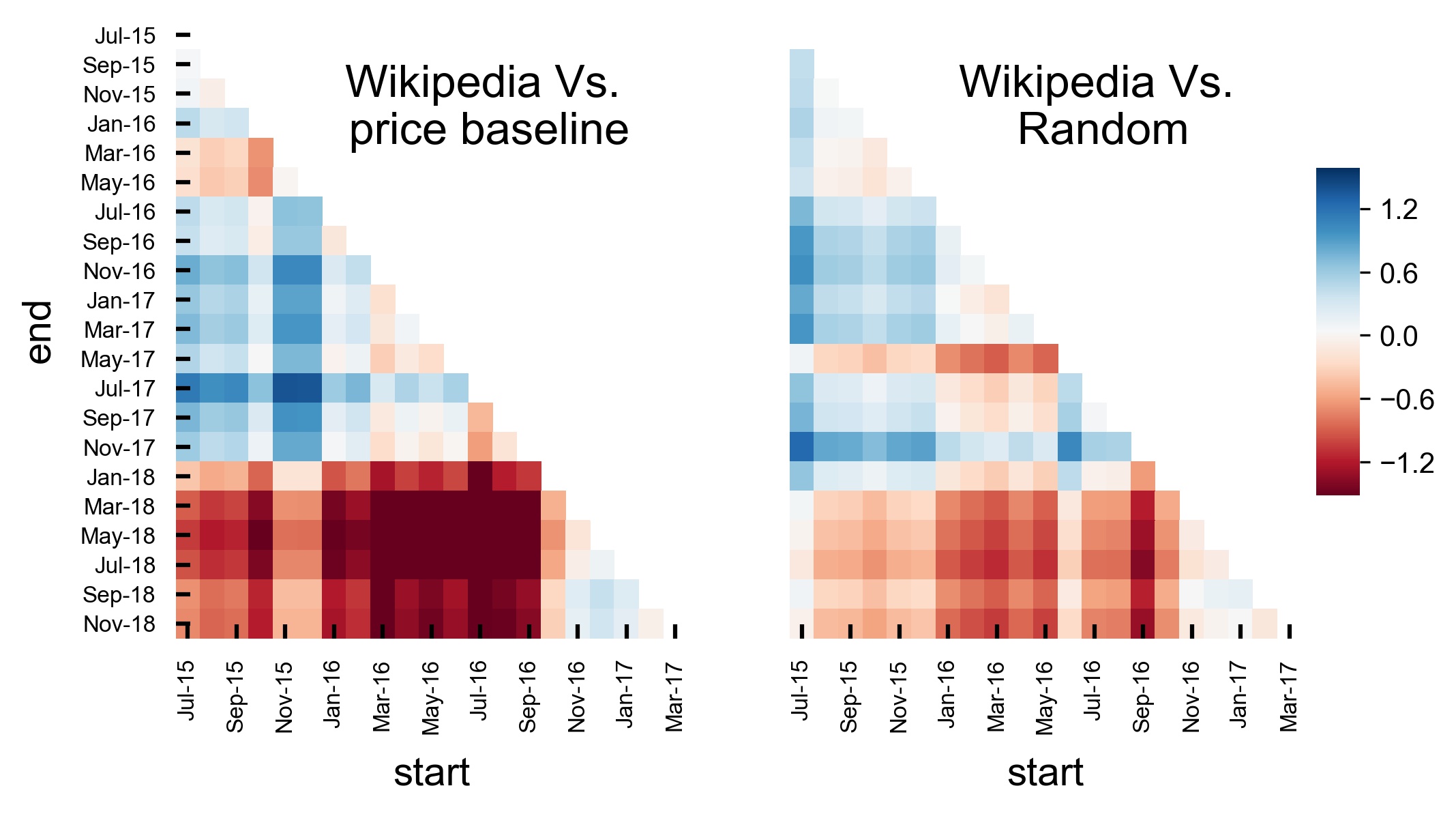}
\caption{\textbf{Comparison between strategies across different periods of time.} Difference between the cumulative log returns of the Wikipedia based strategy and the price based baseline \textbf{(A)} or the random baseline \textbf{(B)} given different start and end dates.}\label{fig:start_end_strateg}
\end{figure}

Finally we investigate the role of the start and end times of the investment period (see Figure \ref{fig:start_end_strateg}). We find that for most of the choices, the Wikipedia-based strategy has a higher cumulative returns than the random strategy. It outperforms the price baseline for the majority of the periods ending before January $2018$. This change after January $2018$ can be attributed to the unexpected turn the market took after Jan $2018$ which caused more than $400$ billion dollars of losses.

\section*{Conclusion and discussion}

In this paper, we have investigated the interplay between the production and consumption of information about digital currencies in Wikipedia and their market performance. We have shown that, over time, there is a positive correlation between the market performance of a cryptocurrency, as measured by its price, volume, and market share and the attention people pay to the corresponding Wikipedia page, measured by the number of page views and the number of page edits. This result suggests that the production and consumption of information in Wikipedia is relevant for investment purposes.

We have analyzed the edit history of cryptocurrency pages in Wikipedia. We have shown that contributions to cryptocurrency pages are bursty in time, with periods of high activity followed by calmer ones. We have found that cryptocurrency pages have experienced a higher number of revert edits ($18\%$) compared to other pages, suggesting they have been subject to vivid debates around their contents. Also, we have found that the number of cryptocurrency pages editors has increased in the period considered, and this is not the case for editors of other topics in Wikipedia. However, very few editors are responsible for most of the edits,  consistently with the rest of Wikipedia. Interestingly, this subset of editors has started contributing relatively recently (after $2012$), also in contrast with the rest of Wikipedia. We have shown that the information in Wikipedia is, to a large extent, provided by cryptocurrency and technology enthusiasts. In fact, we have found that editors who are very active on cryptocurrency pages focus their editing activity almost exclusively on cryptocurrencies and blockchain.
We have found that the community of cryptocurrency editors is tight: On average, each page is connected to $37$ other pages through an average of $7$ editors and active contributors tend to edit many pages. New cryptocurrency pages are typically created by new editors, but then also edited by more experienced ones. For this reason, we find that older pages have higher degree in the co-editing network.

Finally, we have proposed a trading strategy relying on Wikipedia page views and found it yields significant returns compared to baseline strategies, further demonstrating the relevance of Wikipedia for cryptocurrency survival in the market. It is important to mention, however, that our strategy neglects the role played by fees, which could significantly decrease profits in real scenarios. Also, the strategy is not successful since January $2018$, when the cryptocurrency market started suffering major losses.

Characterizing the production and consumption of information around cryptocurrencies is key to
understand the market dynamics and inform investment decisions \cite{de2019fragility}. Although our study was limited to the analysis of Wikipedia data, other sources of information including traditional news outlets , Twitter, Reddit or bitcointalk could reveal important information about the cryptocurrency market dynamics. 

\section*{Acknowledgement}
We would like to thank Miriam Redi from Wikimedia Foundation for her valuable discussion about Wikipedia structure. A.E. acknowledge the support of the Alan Turing Instihttps://www.coindesk.com/tute.

\section*{Data Availability Statement}
The datasets generated and analyzed for this study along with the code to regenerate the figures can be found in \cite{sdata}


\pagebreak
\appendix
\setcounter{figure}{0}
\setcounter{table}{0}
\renewcommand\thefigure{\thesection\arabic{figure}} 
\renewcommand\thetable{\thesection\arabic{figure}} 

\section{Appendix.}

\subsection{List of cryptocurrencies}

We consider for this study all cryptocurrencies with a Wikipedia page. In Table \ref{tab:table_crypt_list}, we present some of their characteristics. Using Wikipedia API, we retrieve data about each page views and edits. For the page views we use the API call: \\ \url{https://wikimedia.org/api/rest_v1/metrics/pageviews/per-article/en.wikipedia.org/all-access/user/wiki_page/daily/start_date/end_date}, \\
where wiki\_page is the cryptocurrency page name and start and end dates are the requested dates. 
To retrieve the edit history, we used the following call: \\
\url{https://en.wikipedia.org/w/api.php?action=query\& format=json\& prop=revisions\& rvprop=timestamp\%7Cuser\%7Ccomment\%7Ccontent&rvlimit=500\&titles=wiki_page}. 

\renewcommand{\arraystretch}{1.5}

\begin{table}[hbtp!]
\centering
\small
\tabcolsep=0.11cm
\caption{\label{tab:table_crypt_list}  \textbf{Cryptocurrencies with a page in Wikipedia.} The table is generated using data collected on January $23$rd, $2019$. The table shows, for the cryptocurrency considered, their name, wikipedia page link, first appearance on the market, page creation date, market capitalization on January $23$rd, $2019$, rank (based on market capitalization), and  whether they can be marginally traded or not. The table is generated using data collected on January $23$rd, $2019$}
\begin{tabular}{@{}*{33}{c}@{}}
\toprule
Name & \thead{Wikipedia page\\link} & \thead{Market entrance\\date} & \thead{Wikipedia page\\creation date } & Market cap (\$) & Rank & Margin trading\\
\midrule
    Auroracoin & {Auroracoin} & {$2014-02-27$} & {$2014-03-16$} & {$1,829,734$} & {$578$} & {No} \\
   	Bitcoin & {Bitcoin} & {$2013-04-28$} & {$2009-03-08$} & {$64,051,590,603$} & {$1$} & {Yes} \\
   	Bitcoin Cash & {Bitcoin\_Cash} & {$2017-07-23$} & {$2017-07-28$} & {$2,165,211,566$} & {$6$} & {Yes} \\
   	Bitcoin Private & {Bitcoin\_Private} & {$2018-03-10$} & {$2018-01-18$} & {$23,125,132$} & {$121$} & {No} \\
   	Bitconnect & {Bitconnect} & {$2017-01-20$} & {$2017-06-28$} & {$6669607$} & {Delisted} & {No} \\
   	Bitcoin Gold & {Bitcoin\_Gold} & {$2017-10-23$} & {$2017-10-15$} & {$183,310,538$} & {$28$} & {Yes} \\
   	Cardano & {Cardano\_(platform)} & {$2017-10-01$} & {$2018-01-10$} & {$1,085,600,679$} & {12} & {Yes} \\
   	Dash & {Dash\_(cryptocurrency)} & {$2014-02-14$} & {$2014-06-01$} & {$688,081,911$} & {$15$} & {Yes} \\
   	Decred & {Decred} & {$2016-02-10$} & {$2017-10-22$} & {$155,712,717	$} & {$32$} & {No} \\
   	Dogecoin & {Dogecoin} & {$2013-12-15$} & {$2013-12-14$} & {$225,747,716$} & {$24$} & {Yes} \\
   	EOS & {EOS.IO} & {$2017-07-01$} & {$2017-11-30$} & {$2,528,271,205$} & {$5$} & {Yes} \\
   	Ethereum & {Ethereum} & {$2015-08-07$} & {$2014-01-27$} & {$12,725,299,480$} & {$2$} & {Yes} \\
   	Ethereum Classic & {Ethereum\_Classic} & {$2016-07-24$} & {$2016-07-25$} & {$436,681,977$} & {$18$} & {Yes} \\
   	Filecoin & {Filecoin} & {$2017-12-13$} & {$2017-08-11$} & {Future} & {$1744$} & {No} \\
   	Gridcoin & {Gridcoin} & {$2015-02-28$} & {$2016-08-30$} & {$1,890,388$} & {$1179$} & {No} \\
   	Litecoin & {Litecoin} & {$2013-04-28$} & {$2012-10-20$} & {$2,594,436,915$} & {$4$} & {Yes} \\
   	MazaCoin & {MazaCoin} & {$2014-02-27$} & {$2014-02-28$} & {$340702$} & {Delisted} & {No} \\
   	Monero & {Monero\_(cryptocurrency)} & {$2014-05-21$} & {$2015-03-19$} & {$814,187,090$} & {$13$} & {Yes} \\
   	Namecoin & {Namecoin} & {$2013-04-28$} & {$2012-06-27$} & {$9,324,709$} & {$242$} & {No} \\
   	NEM & {NEM\_(cryptocurrency)} & {$2015-04-01$} & {$2014-12-11$} & {$353,031,53$} & {$19$} & {No} \\
   	NEO & {NEO\_(cryptocurrency)} & {$2016-09-09$} & {$2017-12-27$} & {$522,866,284	$} & {$16$} & {Yes} \\
   	NuBits & {NuBits} & {$2014-09-24$} & {$2015-11-03$} & {$428,372$} & {$900$} & {No} \\
   	Nxt & {Nxt} & {$2013-12-04$} & {$2014-03-09$} & {$22,864,634$} & {$124$} & {No} \\
   	OmiseGO & {OmiseGO} & {$2017-07-14$} & {$2017-09-14$} & {$162,177,884$} & {$30$} & {Yes} \\
   	Peercoin & {Peercoin} & {$2013-04-28$} & {$2013-04-10$} & {$12,838,440$} & {$188$} & {No} \\
   	Petro & {Petro\_(cryptocurrency)} & {$2014-04-15$} & {$2017-12-03$} & {$969,039	$} & {$1208$} & {No} \\
   	PotCoin & {PotCoin} & {$2014-02-10$} & {$2014-08-06$} & {$4,157,678$} & {$413$} & {No} \\
   	Primecoin & {Primecoin} & {$2013-07-11$} & {$2013-07-29$} & {$3,732,436	$} & {$438$} & {No} \\
   	Ripple & {Ripple\_(payment\_protocol)} & {$2013-08-04$} & {$2005-08-06$} & {$12,531,899,335$} & {$3$} & {Yes} \\
   	Stellar & {Stellar\_(payment\_network)} & {$2014-08-05$} & {$2014-09-04$} & {$1,490,404,115	$} & {$9$} & {Yes} \\
\bottomrule
\end{tabular}
\end{table}

\begin{table}[hbtp!]
\centering
\small
\tabcolsep=0.11cm
\begin{tabular}{@{}*{8}{c}@{}}
\toprule
Name & \thead{Wikipedia page\\link} & \thead{Market entrance\\date} & \thead{Wikipedia page\\creation date } & Market cap (\$) & Rank & Margin trading\\
\midrule
   	Tether & {Tether\_(cryptocurrency)} & {$2015-02-25$} & {$2017-12-05$} & {$2,031,040,167$} & {$7$} & {Yes} \\
   	Tezos & {Tezos} & {$2017-10-02$} & {$2018-06-28$} & {$233,324,546$} & {$23$} & {No} \\
   	Titcoin & {Titcoin} & {$2014-08-26$} & {$2014-09-13$} & {$20,798$} & {$1602$} & {No} \\
   	Verge & {Verge\_(cryptocurrency)} & {$2014-10-25$} & {$2018-01-24$} & {$90,702,804$} & {$49$} & {No} \\
   	Vertcoin & {Vertcoin} & {$2014-01-20$} & {$2014-03-12$} & {$14,488,303$} & {$175$} & {No} \\
   	Waves platform & {Waves\_platform} & {$2016-06-02$} & {$2017-04-27$} & {$276,964,620$} & {$21$} & {No} \\
   	Zcash & {Zcash} & {$2016-10-29$} & {$2016-10-03$} & {$299,380,368$} & {$20$} & {Yes} \\  	
\bottomrule
\end{tabular}
\end{table}

\subsection{Exchanges with margin trading support}
\label{margin_trade}

Here, we provide data on the list of exchanges supporting margin trading. Margin trading is essential for our proposed investment strategy, since an investor can sell a cryptocurrencies which he does not own yet. 

\begin{table}[hbtp!]
\centering
\small
\tabcolsep=0.11cm
\caption{\label{tab:table_exchange_list}  \textbf{List of exchanges supporting margin trading} The table is generated using data collected on January $23$rd, $2019$. It shows the names and webpage urls of the exchanges considered.}
\begin{tabular}{@{}*{33}{c}@{}}
\toprule
Name & Link \\
\midrule
    Bitmax & {\url{https://www.bitmex.com} }  \\
    Huobi &{\url{https://www.huobi.co}} \\
    poloniex & {\url{https://poloniex.com}} \\
    kraken &  \url{https://www.kraken.com}\\
    Bitfinex & \url{https://www.bitfinex.com}\\
   \bottomrule
\end{tabular}
\end{table}

\subsection{Correlations between Wikipedia page views and market properties.}
\label{ap_corr}
\begin{figure}[hbtp!]
  
\centering\includegraphics[scale=0.8]{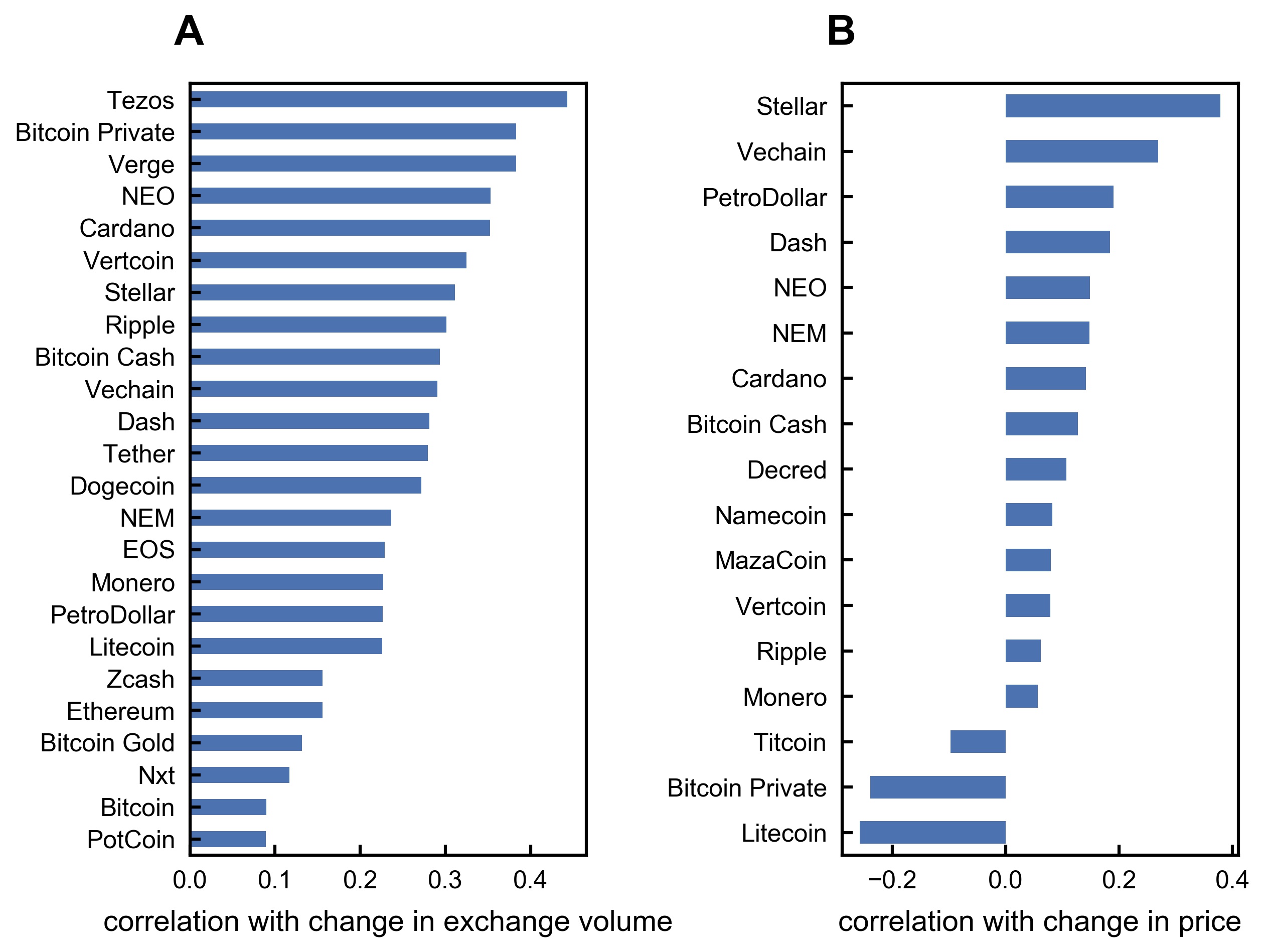}
\caption{\textbf{Correlation between market properties and Wikipedia page views.} The Pearson correlation between changes in Wikipedia page views and changes in trading volume (A) or price (B), measured across a cryptocurrency entire existence. Only significant correlations ($p<0.05$) are shown.}\label{fig:ap_wiki_views_mark_perf_crypt}
\end{figure}

\begin{figure}[hbtp!]
  
\centering\includegraphics[scale=0.8]{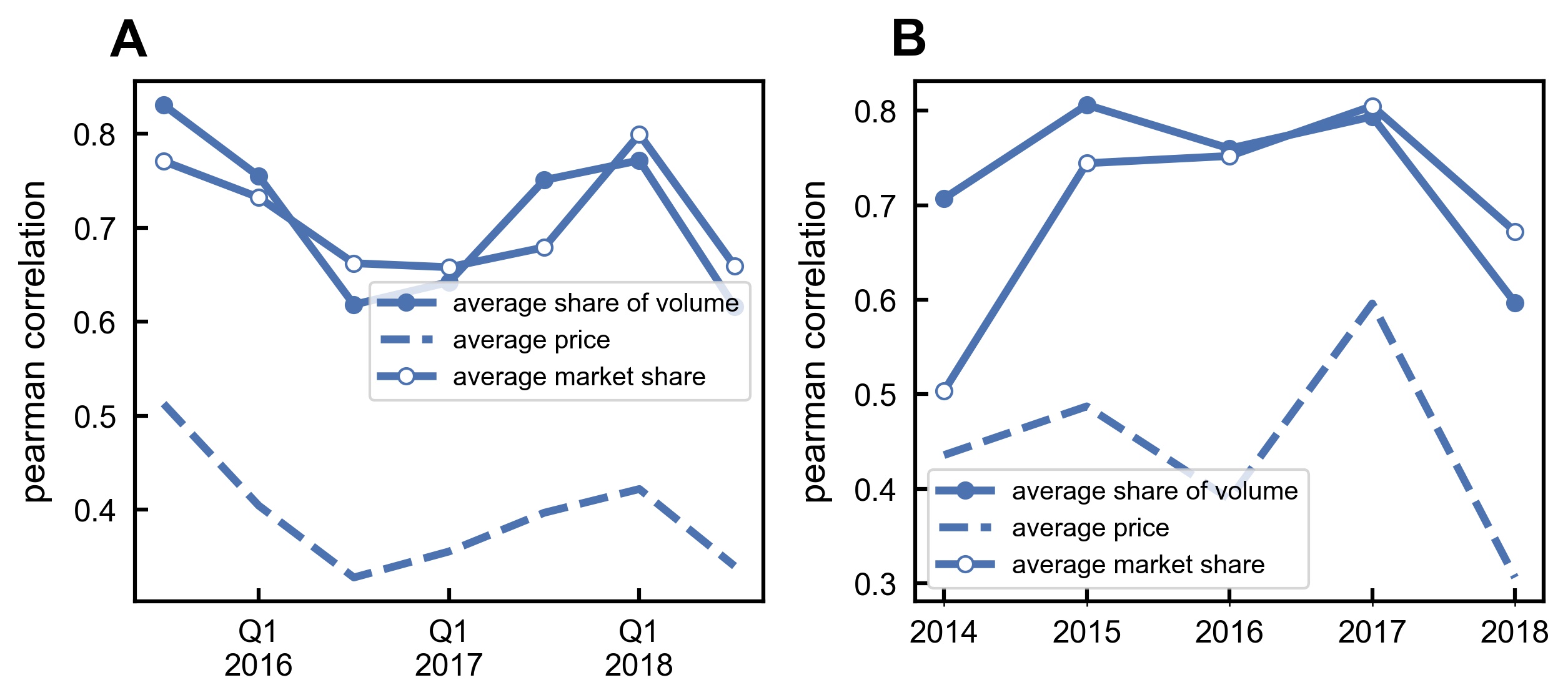}
\caption{\textbf{Persistency of the correlation between market properties and attention on Wikipedia.} (A) The Spearman correlation between average share of views and price (blue dashed line),   volume (blue line with filled circles) and market share (blue line with white circles) across time. The correlation is computed over a window of $6$ months. (B) The Spearman correlation between a cryptocurrency page average share of edits and price (blue dashed line), volume (blue line with filled circles) and market share (blue line with white circles) across time. The correlation is computed over a window of one year.}\label{fig:ap_wiki_market_corr_time}
\end{figure}

The number of Wikipedia page views and the properties of the market are overall correlated. In Figure \ref{fig:ap_wiki_views_mark_perf_crypt}, we show the correlations between Wikipedia page views, trading volume and price for the cryptocurrencies considered. We show the Spearman correlation between a cryptocurrency average share of page views and the market performance measured by its average market share ($\rho_{vm}$),  average trading volume share ($\rho_{vv}$) and average price ($\rho_{vp}$) across time (see Figure \ref{fig:ap_wiki_market_corr_time}A). We show that the positive correlation between this quantities is consistent with time, with $ 0.65\leq\rho_{vm}\leq0.79$,  $0.61\leq\rho_{vv}\leq0.83$, and $0.32\leq\rho_{vp}\leq0.51$.

In Figure \ref{fig:ap_wiki_market_corr_time}-B, we show the Spearman correlation between a cryptocurrecny average share of Wikipedia page edits and its market performance measured in average market share ($\rho_{em}$),  average trading volume share ($\rho_{ev}$) and average price ($\rho_{ep}$) across time. We show that the positive correlation between this quantities is consistent with time, with $ 0.25\leq\rho_{em}\leq0.78$,  $0.21\leq\rho_{ev}\leq0.79$, and $0.32\leq\rho_{ep}\leq0.66$. However the value of the correlation varies across the years which can be attributed to the variation in the number of pages created per year.

\subsection{Literature review.}
Several studies have focused on Wikipedia pages and editors' activity. In Table \ref{tab:lit_sum}, we present a summary of their findings and a comparison with our results around cryptocurrencies Wikipedia pages.

\begin{table}[hbtp!]
\caption{\label{tab:lit_sum}  \textbf{Comparison among our results and previous findings around Wikipedia pages and editors.} The table reports for each research paper: (1) Reference. (2) Focus of the article. (3) Key measurements. (4) Key findings relevant to our study. (5) Our findings around cryptocurrency pages in comparison to the previous findings. }
\begin{tabularx}{\linewidth}{c L L L L}
\toprule
Paper reference & Focus & Key measurements & Findings & Our findings \\
\midrule
\cite{kittur2007he} (kittur) & {editors} & {Fraction of maintenance 
edits.} & General increase in maintenance work, especially reverts. & Higher proportion of reverts. No increasing trend in both reverts and vandalism.\\ 
\midrule
\cite{panciera2009wikipedians} (panciera) & {editors} & {Editors activity levels in relation to their life time} & {Highly active editors (Wikipedians) are active from two days after joining Wikpedia.} & {Similar findings for cryptocurrency pages (see Figure \ref{fig:edits_activity})}  \\
\midrule
\cite{kittur2007power} (kittur) & {editors} & {Evolution of the contributions of editors given their activity levels.} & {Growth in the number of infrequent contributors and increase in their number of edits.} & {Infrequent editors have existed since the beginning and their number of edits also increases (see Figures \ref{fig:editors_activity} and \ref{fig:edits_activity})}. \\

\midrule
\cite{heilman2015wikipedia} (heilman) & {Medical related Wikipedia pages} & {Descriptive analysis of the general trends.} & {Decreasing number of editors} & {Increasing number of editors (see Figure \ref{fig:views_char}).}  \\
\bottomrule
\end{tabularx}
\label{table:comp}
\end{table}

\subsection{Robustness of the findings.}
\label{ap:robust}
\begin{figure}[hbtp!]  
\centering\includegraphics[scale=0.8]{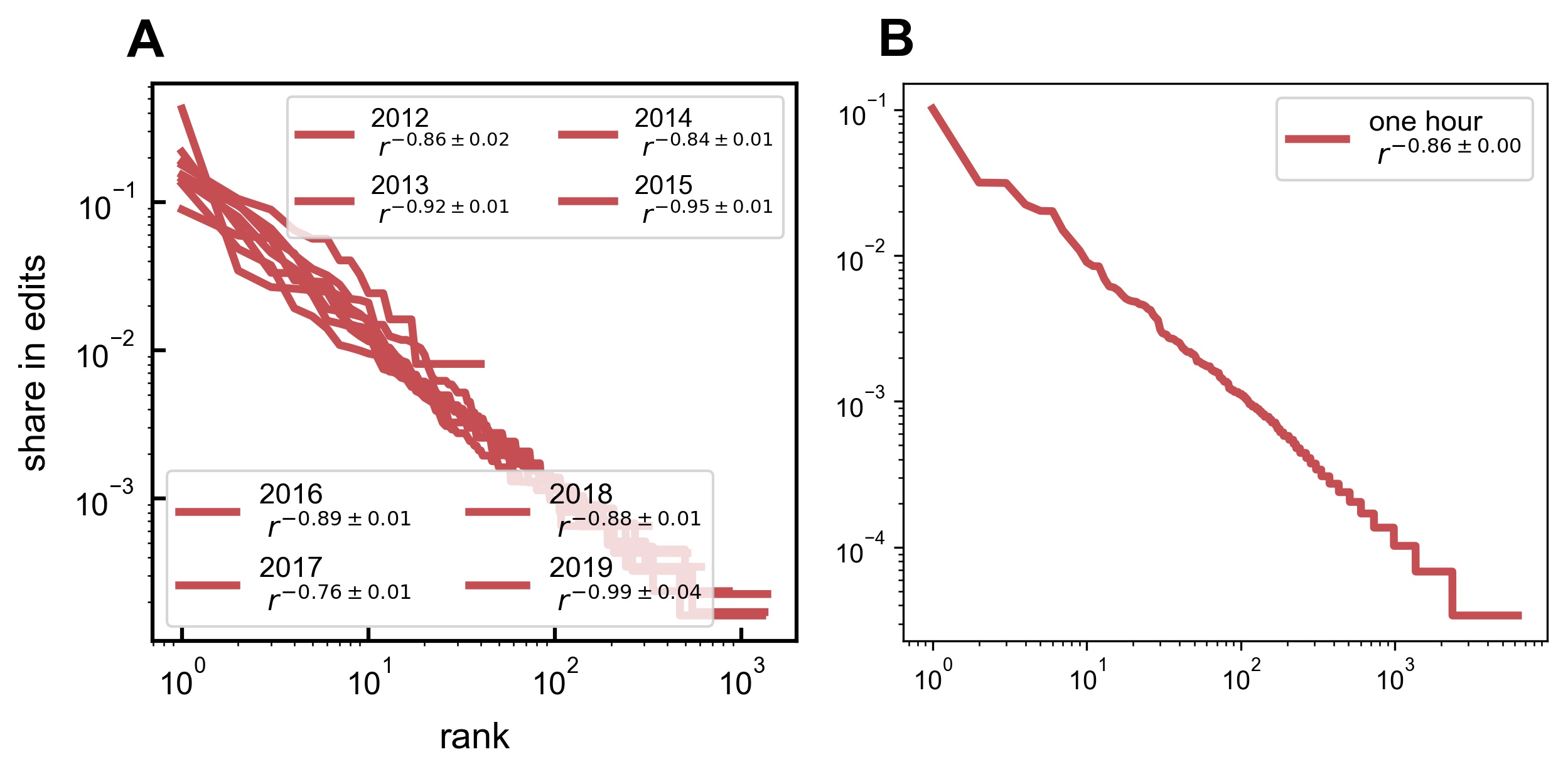}
\caption{\textbf{Users share of edits in different years} (A) The fraction of edits vs the rank $r$ of an editor, computed over a year. Every line represents different year. (B) The fraction of edits vs the rank $r$ of an editor, computed over the period between $2005$ and $2018$, after removing edits from the same editor on the same page, occurring within one hour from the previous.}\label{fig:ap_edits_share_years}
\end{figure}

The uneven distribution of edits across editors was depicted in Figure \ref{fig:editors_share}. Here, we show that this result is consistent in time (see Figure \ref{fig:ap_edits_share_years}-A). We also test our results against saving mistakes by editors \cite{panciera2009wikipedians}. This often occurs when an editor mistakenly save an incomplete edit, producing multiple edits within a very short time. We solve this issue by excluding from the analysis edits that from the same editor on the same page, occurring within less than an hour from the prevopus one, as in \cite{panciera2009wikipedians}. In Figure \ref{fig:ap_edits_share_years}-B, we show that, our results are robust to this change and well described by a power-law distribution $(P(r)\sim r^{-\beta})$ with exponent $\beta = 0.62$.

\begin{figure}[hbtp!]  
\centering\includegraphics[scale=0.8]{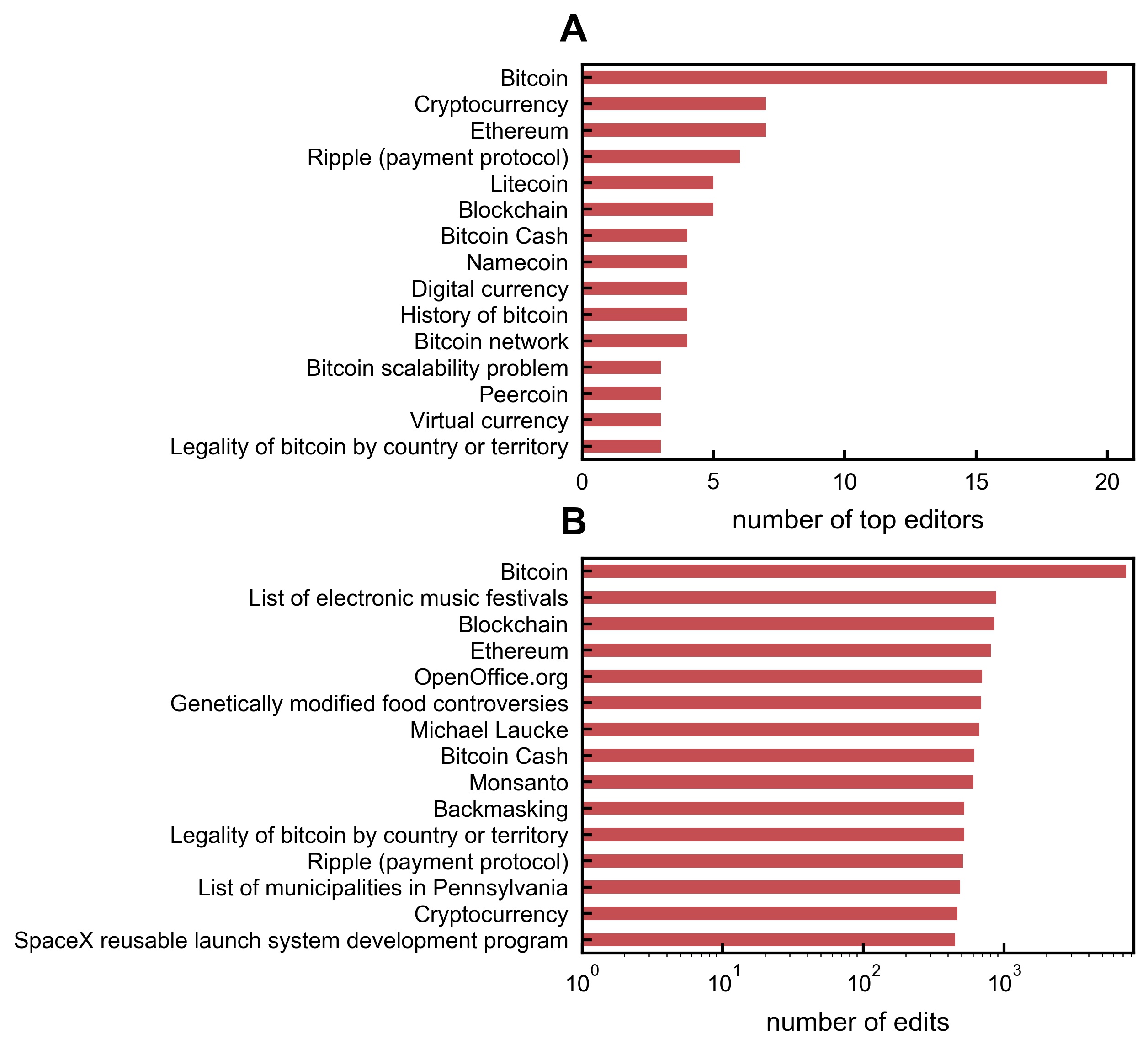}
\caption{\textbf{Activity of the top cryptocurrency pages editors.}(A) The top 15 pages by number of editors. The x-axis shows the number of top editors who had this page in their top edited pages. Note that here we consider only the top 10 pages per editor. (B) The top 15 pages by number of edits. The x-axis shows the total number of edits per page. Results are obtained for the subset of $23$ most active editors.}\label{fig:ap_top_ediotrs_out}
\end{figure}

We also study top editors contributions in all Wikipedia pages. For each editor with at least $100$ edits in cryptocurrency pages, we collect data about the top $10$ Wikipedia pages they contributed. This include pages outside the $38$ cryptocurrency pages. For this task, we use a web tool \cite{xtool}, which provides the number of edits contributed by each editor to a given page. Figure \ref{fig:ap_top_ediotrs_out} shows that editors are mostly interested in cryptocurrencies and technology related pages. Compared to the set editors with more than $500$ edits (see Figure \ref{fig:top_edit_out}), the set of pages edited is more diverse.

\subsection{New pages}
\label{ap:new_pages}

\begin{figure}[hbtp!]  
\centering\includegraphics[scale=0.8]{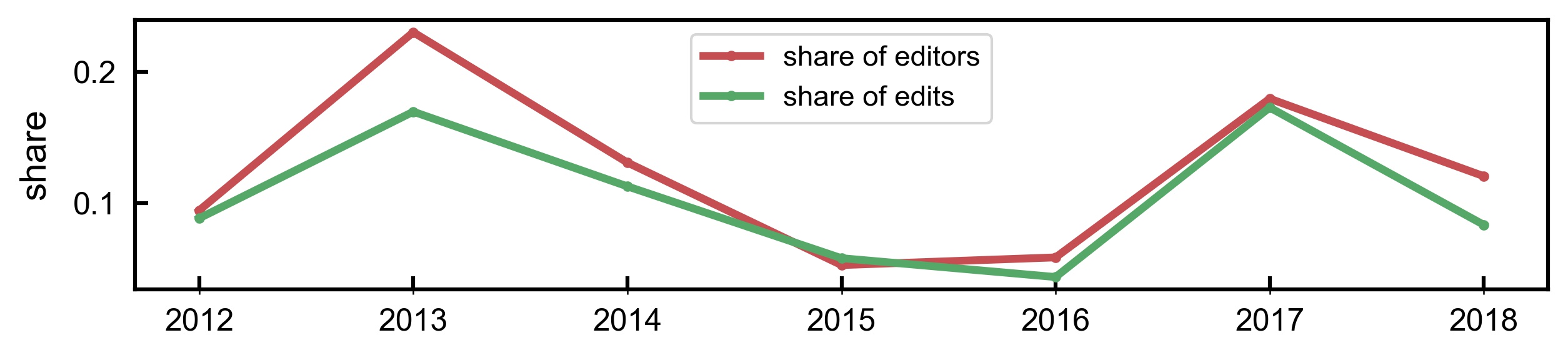}
\caption{\textbf{Editing activity on new pages.} Fraction of edits made to new pages (green solid line), and fraction of editors contributing to new pages (red solid line). Results are aggregated using a time window of one year.}\label{fig:new_pages}
\end{figure}

Figure \ref{fig:new_pages} shows, for each of the years considered, the fraction of edits made to new pages and the fraction of editors contributing to new pages. On average, the $\sim 18\%$ of editors contribute to the newly created pages within a given year, while only $\sim 10\%$ of the edits are made to new pages.

\subsection{The most active editor}
\label{ap:very_active}

Here, we provide information on the editor with the highest number of edits in cryptocurrency pages ($10\%$ of the edits). Table \ref{tab:top_editor} shows the editor general editing patterns in the entire English Wikipedia. Table \ref{tab:top_top_pages} shows the top pages edited by the top editor.

\begin{table}[hbtp!]
\centering
\small
\tabcolsep=0.11cm
\caption{\label{tab:top_editor}  \textbf{Overall activity of cryptocurrency Wikipedia pages top editor.} The table reports for the editor with highest contribution in cryptocurrencies Wikipedia page ($\sim 10\%$ of the edits): (1) number of pages edited, (2) total number of edits (English Wikipedia), (3) percentage of edits in cryptocurrency pages, (4) average number of edits per page, (5) date of the first edit. These data cover the editor activity across all pages in Wikipedia and it was collected on February $5$th, $2019$.}
\begin{tabular}{@{}*{33}{c}@{}}
\toprule
number of page & total number of edits & \thead{percentage of edits\\ in cryptocurrency pages} &average edits per page & date of first edit \\
\midrule
$442$ & $9430$ & $32\%$ & $2$ & $2005-11-20$    \\
\bottomrule
\end{tabular}
\label{table:top_editor}
\end{table}

\begin{table}[hbtp!]
\centering
\small
\tabcolsep=0.11cm
\caption{\label{tab:top_top_pages}  \textbf{Top pages edited by the top editor} The top pages edited by the most active editor in cryptocurrency pages. The table shows the page name and number of edits. Data was collected on February $5$th, $2019$.}
\begin{tabular}{@{}*{33}{c}@{}}
\toprule
page name & number of edits \\
\midrule
    Bitcoin & $2706$  \\
    Blockchain & $593$\\
    Legality of bitcoin by country or territory & $467$ \\
    Bitcoin Cash &  $349$ \\
    Cryptocurrency & $308$\\
    Rebol &$209$\\
    Bitcoin scalability problem & $187$\\
    History of bitcoin  & $177$\\
    Satoshi Nakamoto & $109$\\
   \bottomrule
\end{tabular}
\end{table}

\subsection{Editing network}
\begin{figure}[H]
  
\centering\includegraphics[scale=0.8]{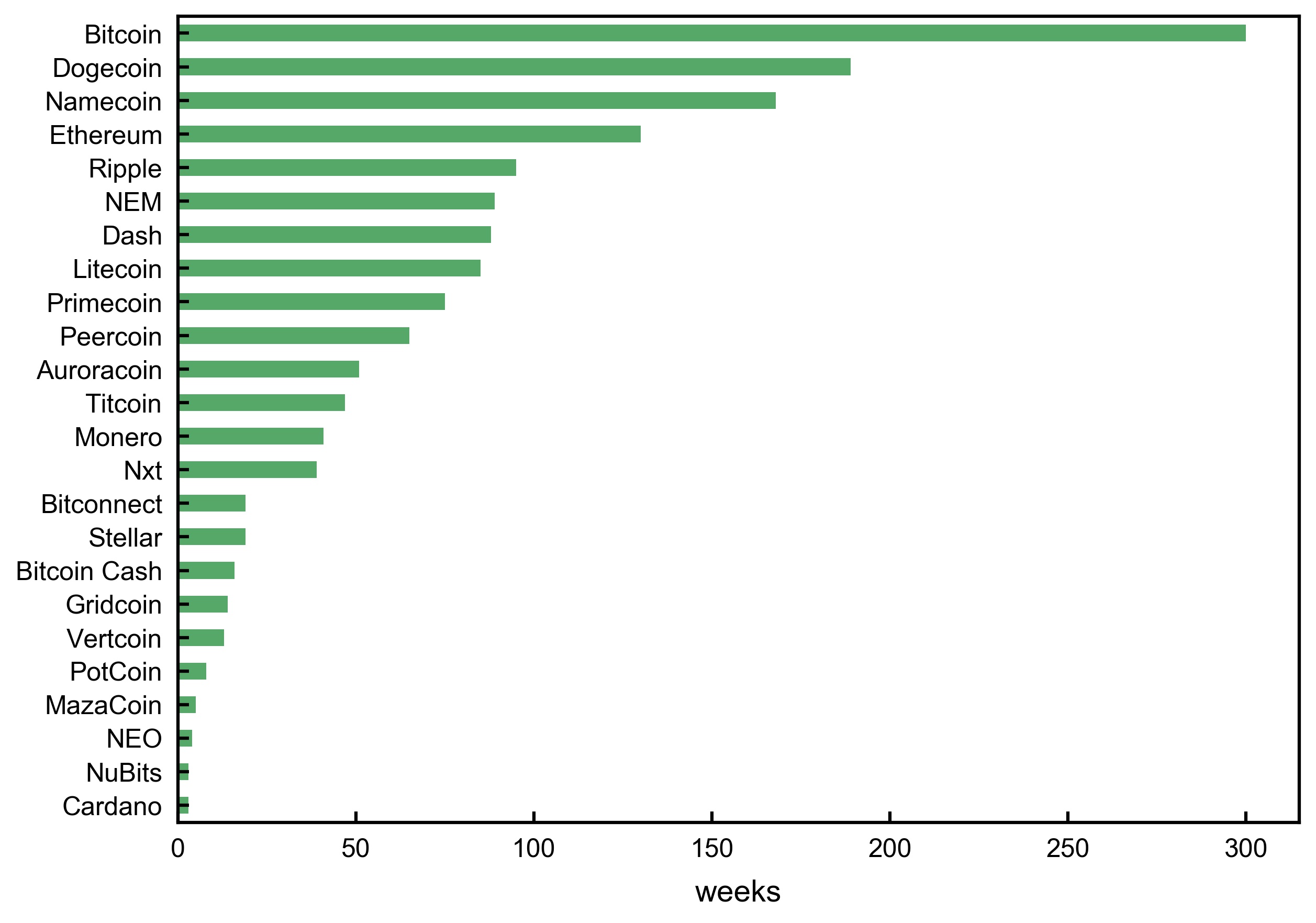}
\caption{\textbf{Ranking in degree centrality.} Number of weeks a cryptocurrency occupied one of top $5$ ranks based on degree centrality in the co-editing Wikipedia pages network. }\label{fig:ap_cent_years}
\end{figure}

To characterize the co-editing activity in cryptocurrency Wikipedia pages, we constructed a weighted undirected network. A node represents a Wikipedia page and an edge exists between two nodes if they have at least one editor in common. Weights on edges represent the number of editors in common. We look at the evolution of the network across time and identify the most central pages according to the degree centrality. Figure \ref{fig:ap_cent_years} shows the number of weeks each cryptocurrencies appeared in the top $5$ ranks when cryptocurrencies are ranked according to their degree centrality in descending order.

\begin{figure}[H]
  
\centering\includegraphics[scale=0.8]{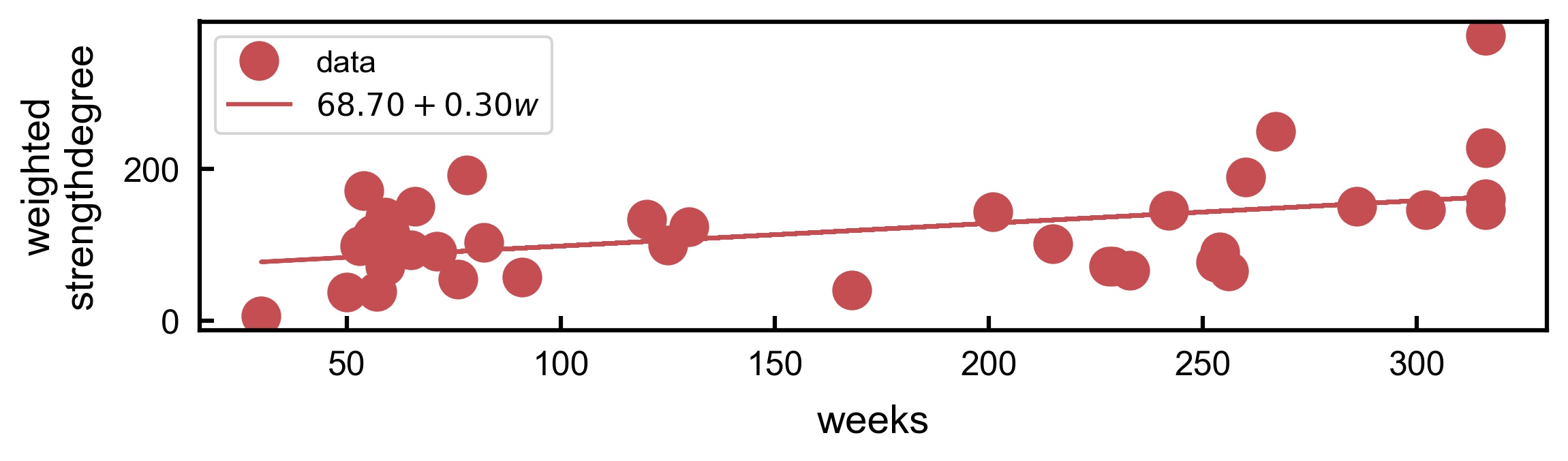}
\caption{\textbf{Correlation between page age and network strength.} Page age in weeks vs its weighted degree in the editing network. Each point represents a node (page). Pearson correlation $\rho = 0.40, p = 0.015$. The solid line represents a fit $a+bw$ where $b = 0.28 \pm 0.10$. }\label{fig:ap_s_t}
\end{figure}

Figure \ref{fig:ap_s_t} shows the correlation between the age of a cryptocurrency page and its weighted degree ($\rho = 0.40, p = 0.015$).

\end{document}